\documentclass[universe,article,accept,pdftex,oneauthor]{Definitions/mdpi} 
\firstpage{1} 
\pubvolume{1}
\issuenum{1}
\articlenumber{0}
\pubyear{2025}
\copyrightyear{2025}
\externaleditor{Firstname Lastname} 
\datereceived{18 March 2025} 
\daterevised{19 April 2025} 
\dateaccepted{21 April 2025 } 
\datepublished{ } 
\hreflink{\textls[-15]{https://doi.org/}} 
\Title{Reconciling the Waiting Time Peaks Variations of Repeating FRBs with an Eccentric Neutron Star--White Dwarf Binary}

\TitleCitation{Reconciling the Waiting Time Peaks Variations of Repeating FRBs with an Eccentric Neutron Star--White Dwarf Binary}

\Author{{Hao-Yan Chen} 
 \orcidA{}}

\AuthorNames{Hao-Yan Chen}

\AuthorCitation{Chen, H.-Y.}

\address[1]{School of Science, Huzhou University, {Huzhou} 
 313000, China; chenhaoyan@zjhu.edu.cn
}

\abstract{Fast radio bursts (FRBs) are luminous radio transients with millisecond duration. For some active repeaters, such as FRBs 20121102A and 20201124A, more than a thousand bursts have been detected by the Five-hundred-meter Aperture Spherical radio Telescope (FAST). The waiting time (WT) distributions of both repeaters, defined as the time intervals between adjacent (detected) bursts, exhibit a bimodal structure well-fitted by two log-normal functions. Notably, the time scales of the long-duration WT peaks for both repeaters show a decreasing trend over time. These similar burst features suggest that there may be a common physical mechanism for FRBs~20121102A and 20201124A. In this paper, we {revisit} the neutron star (NS)--white dwarf (WD) binary model with an eccentric orbit to account for the observed changes in the long-duration WT peaks. According to our model, the shortening of the WT peaks corresponds to the orbital period decay of the NS-WD binary. We consider two mass transfer modes, namely, stable and unstable mass transfer, to examine how the orbital period evolves. Our findings reveal distinct evolutionary pathways for the two repeaters: for FRB~20121102A, the NS-WD binary likely undergoes a combination of common envelope (CE) ejection and Roche lobe overflow, whereas for FRB~20201124A the system may experience multiple CE ejections. These findings warrant further validation through follow-up observations.}


\begin{document}

\section{Introduction} \label{sec:intro}

Fast radio bursts (hereafter, FRBs) are luminous transients with millisecond duration (see reviews, {e.g.,} 
~\citep{2019ARA&A..57..417C,Petroff et al.(2019),Petroff et al.(2022)}). Because the dispersion measures of FRBs exceed the contributions of our Galaxy, and certain FRB host galaxies have been identified at redshifts in the range $z \simeq 0.03-1$ (e.g.,~\citep{Marcote et al.(2020),Ryder et al.(2023)}), FRBs are thought to be extragalactic origins (except FRB~20200428 detected from the Galactic magnetar SGR J1935+2154;~\citep{Bochenek et al.(2020),CHIME/FRB Collaboration et al.(2020b)}). Based on the quantity of detected bursts, FRBs can be classified as repeating FRBs and apparently one-off (hereafter, simply ``one-off'') FRBs.
Recently, nearly 500 unique FRB sources have been detected by {the Canadian Hydrogen Intensity Mapping Experiment (CHIME)}
(i.e., the CHIME/FRB Catalog {1} 
\endnote{\url{https://www.chime-frb.ca/catalog} {(accessed on 22 April 2025).} 
~\citep{Amiri et al.(2021)})}. This sample is excellent for exploring the statistical properties of FRBs. Some statistical differences between one-off FRBs and repeating ones have been found based on this sample, including the differences in the distributions of the peak fluxes 
\citep{Amiri et al.(2021)}, the spectral indices~\citep{Zhong et al.(2022)}, and the luminosities~\citep{Chen et al.(2022b),Cui et al.(2022)}. Additionally, the time--frequency drift (``sad-trombone'' effect) commonly seen in repeating FRBs (e.g.,~\citep{Hessels et al.(2019)}) is mostly absent in one-offs. All of these differences strongly suggest that one-off and repeating FRBs likely have distinct emission mechanisms and progenitors (e.g.,~\citep{Zhang(2020),Zhang(2023)}).

Periodic activities are an interesting property of repeaters that can
provide important clues for revealing their physical mechanisms. Analysis of the burst arrival times of 
FRB~20180916B shows a $\sim$$5$-day activity window\endnote{The length of the activity window depends on observing frequency. For example, the activity phase of the bursts detected by LOFAR in the 110--188~MHz is $\sim$$ 25\%$, whereas the activity window of bursts detected by CHIME/FRB (400--800 MHz) is $\sim$$ 30\%$ of the activity cycle~\citep{Gopinath et al.(2024)}.}
that recurs every 16.35 days~\citep{CHIME/FRB Collaboration et al.(2020a)}.
FRB 20121102A, the first detected repeater coincident with a compact persistent radio source~\citep{Chatterjee et al.(2017),Marcote et al.(2017)}, was found to exhibit a candidate periodicity of $\sim$$ 160$ days~\citep{Rajwade et al.(2020),Cruces et al.(2021)}. 
Some models have been proposed to explain these long
activity periods, including orbital motion (e.g.,~\citep{2020ApJ...895L...1D,Gu et al.(2020),Lyutikov et al.(2020),Deng et al.(2021),Chen et al.(2022a),Lin et al.(2022)}), rotational period (e.g.,~\citep{Beniamini et al.(2020)}),
and precession-based mechanisms (e.g.,~\citep{Levin et al.(2020),2020ApJ...893L..31Y,
Chen et al.(2021),Sridhar et al.(2021)}).

The Five-hundred-meter Aperture Spherical radio Telescope (FAST) has conducted continuous monitoring of
several active repeating FRBs, including FRBs~20121102A and 20201124A
(e.g.,~\citep{Li et al.(2021), Xu et al.(2022),Zhang et al.(2022)}). 
Over the course of 59.5 h of observation time between 29 August 2019 and 29 October 2019, 
\citet{Li et al.(2021)} detected 1652 independent burst events from FRB 20121102A.
These bursts have a bimodal energy distribution that can be well-fitted by combining a log-normal function with a Cauchy function
\citep{Li et al.(2021)}. Moreover, the distribution of the waiting time, defined as the time interval between two adjacent (detected) bursts, demonstrates a bimodal structure. The two fitted peak waiting times are $\sim$$3.4$ ms and $\sim$$70$ s 
(the fitted peak of the waiting time distribution of the high-energy 
component ($E> 10^{38} \ \rm{erg}$) is $\sim$$220$ s;~\citep{Li et al.(2021)}). 
For FRB~20201124A\endnote{This repeater was first discovered by
CHIME/FRB~\citep{Lanman et al.(2022)}.}, 1863 separate burst have been detected by FAST
during a total observation time of 82 h from 1 April 2021 to 29 May 2021~\citep{Xu et al.(2022)}. 
The waiting time distribution of FRB 20201124A similarly reveals a bimodal structure, which can be well-fitted by two log-normal functions
peaking at 39 ms and 106.7 s, respectively~\citep{Xu et al.(2022)}. 
Despite the detection of thousands of bursts, no significant periods have been found in the millisecond to the second range for either FRBs~20121102A or 20201124A (e.g.,~\citep{Li et al.(2021),Niu et al.(2022),Xu et al.(2022)}).

For FRB~20121102A, a total of 478 independent burst events were 
detected by the \mbox{305-m} Arecibo telescope\endnote{{The detection threshold for Arecibo is different from that of FAST, i.e., 0.015 Jy ms for FAST~\citep{Li et al.(2021)} and 0.057 Jy ms for Arecibo~\citep{Hewitt et al.(2022)}.}} over 59 h of observation between
December 2015 and October 2016~\citep{Hewitt et al.(2022)}.
The waiting time distribution of these bursts similarly exhibits a bimodal structure, which can be fitted by 
two log-normal functions peaking at $\sim$$24$ ms and $\sim$$95$ s, respectively~\citep{Hewitt et al.(2022)}. 
For FRB 20201124A, following the detection of a new burst by
CHIME/FRB on 21 September 2021\endnote{\url{https://www.chime-frb.ca/repeaters/
FRB20201124A} {(accessed on 22 April 2025).}
}, \citet{Zhang et al.(2022)} used FAST to monitor this source,
initiating observations on 25 September 2021. In all, 881\endnote{The number of bursts from FRB 20201124A 
detected by FAST is distinct among the four papers in this series,
i.e.,~\citep{Jiang et al.(2022),Niu et al.(2022),Zhang et al.(2022),Zhou et al.(2022)}
as distinct criteria have been adopted by the different groups for different
scientific purposes.} independent burst events from FRB 20201124A were 
detected in the first four days (i.e., from 25 September 2021 to 28 September 2021;~\citep{Zhang et al.(2022)})
of the 17-day monitoring campaign.
The waiting time distribution of these bursts can also be fitted by two 
log-normal functions peaking at 51.22 ms and 10.05~s, respectively 
\citep{Zhang et al.(2022)}. The time scales of the long-duration (second-level) peaks of the 
waiting time distributions of these two active repeaters both decrease
with time, which would be a sign that both FRB sources are becoming increasingly active.

These observed similarities in burst characteristics suggest that a common physical mechanism likely exists for both FRBs~20121102A and 20201124A. {In this paper, we explain the common features observed in two repeaters using the NS-WD binary model with an eccentric orbit~\citep{Gu et al.(2020)}. We consider that the time intervals between adjacent bursts correspond to the orbital period of the NS-WD binary, as the intermittent type of Roche lobe overflow can commonly occur in the semi-detached systems. Consequently, the {decreases} in the long-duration wait time peaks are linked to the decreases in the orbital periods of the NS-WD binary. To investigate the evolution of the NS-WD binary, we consider two distinct forms of mass transfer, namely, stable and unstable mass transfer.}

The remainder of this article is organized as follows: the NS-WD binary model with an eccentric orbit is illustrated in Section~\ref{sec:model}; the evolution of the NS-WD binary with stable mass transfer is studied in Section~\ref{sec:stable}; the evolution of the binary with a massive WD based on a form of unstable mass transfer is shown in Section~\ref{sec:unstable}; finally, conclusions and discussion are presented in Section~\ref{sec:con}.

\section{NS-WD Binary Model} \label{sec:model}

\citet{Gu et al.(2020)} proposed a compact binary model with an eccentric orbit to explain the periodic activity observed in repeating FRBs. 
The system contains a magnetic WD and an NS with strong dipolar magnetic fields.
At the periastron, the mass transfer occurs from the WD to the NS through the inner Lagrange point $L_{1}$ when the WD fills its Roche lobe. At other positions along the eccentric orbit, no mass transfer is supplied, as the Roche lobe is not filled.
When the accreted magnetized materials approach the NS surface, magnetic reconnection is triggered, releasing strong electromagnetic radiation via curvature radiation as the electrons move along the NS magnetic field lines.
The WD may be kicked away after a Roche lobe overflow process if $q < 2/3$~\citep{King(2007)}, where $q$ is the mass ratio defined as $q=M_{\rm{WD}}/M_{\rm{NS}}$, with $M_{\rm{NS}}$ and $M_{\rm{WD}}$ respectively denoting the masses of the NS and WD. The WD can replenish its Roche lobe through gravitational radiation; therefore, the next transfer process can recur. The intermittent nature of Roche lobe overflow appears to be a common phenomenon in the semi-detached NS-WD system, as the NS mass is normally larger than $1.4 \ M_{\odot}$ and the WD mass distribution peaks at $0.6 \ M_{\odot}$. According to this model, the duty cycle of the burst activity is related to the orbital period of the binary. An extremely high eccentricity ($e > 0.95$) of orbit is required to explain the 16.35-day periodicity of FRB 20180916B~\citep{Gu et al.(2020)}.

The wait time distributions of the active repeaters FRBs~20121102A and 20201124A both exhibit a bimodal structure, i.e., a millisecond peak and a tens-of-second peak. A possible explanation for the millisecond peak is that due to the high time resolution of the radio telescopes, e.g., the~98.304-$\mu s$ sampling resolution of FAST~\citep{Li et al.(2021)}, the individual burst events with long-duration pulse widths and multi-component pulse profiles (e.g.,~FRB~20191221A including 3-second duration and nine components;~\citep{CHIME/FRB Collaboration et al.(2022)}) can be considered to be multiple separate bursts. Following the spirit of \citet{Gu et al.(2020)}, these bursts with multiple pulse components can be produced in the mass transfer process. Specifically, the viscous processes during mass transfer are necessary to help the accreted materials lose angular momentum and eventually fall onto the surface of the NS. During this process, some massive accreted materials may be fragmented into a number of smaller portions. When the fragmented materials originating from the same initial bulk of material approach the NS at different times, multiple bursts with short intervals may be triggered, which can be regarded as a single burst event containing multiple sub-bursts.

For the tens-of-second peaks, an intriguing phenomenon is that the long-duration peaks decay with time. In our hypothesis, these peak waiting times {may correspond to} the orbital periods $P_{\rm{orb}}$ of the NS-WD binary. As a result, the variations in the orbital periods of the NS-WD binary induced by the mass transfer are hypothesized to induce variations in the long-duration wait time peaks of repeating FRBs.
The reasons for this are as follows. For the semi-detached NS-WD system, the intermittent type of Roche lobe overflow may be a common phenomenon (e.g.,~\citep{King(2007),Gu et al.(2016),Gu et al.(2020)}). The time interval between adjacent mass transfer processes $T_{\rm{mt}}$ is proportional to the mass of the transferred material from the WD $\Delta M_{\rm{WD}}$~\citep{Lin et al.(2022)}. For $M_{\rm{WD}} \sim 0.6\ M_{\odot}$, it can be evaluated that $T_{\rm{mt}} \lesssim P_{\rm{orb}}$ (see Equation (9) of~\citep{Lin et al.(2022)}). Moreover, the supply of the accreted materials to the NS is expected to exhibit a uniform temporal lag. Therefore, the time interval between two adjacent bursts should be equivalent to $P_{\rm{orb}}$.

On the other hand, due to the remaining angular momentum, some of the fragmented materials cannot not fall to the surface of the NS while surrounding it; that is, even though no mass transfer occurs in the binary system, some materials surrounding the NS arrive on its surface episodically, where they stochastically produce radio bursts with wait times ranging from a few seconds to thousands of seconds.

\section{Stable Mass Transfer} \label{sec:stable}

\subsection{Orbital Period} \label{sec:period}

The dynamic equation of the NS-WD binary system is provided {by} 
\begin{equation}
\frac{G \left(M_{\rm{NS}}+M_{\rm{WD}} \right)}{a^{3}}=\frac{4 \pi^{2}}
{P_{\rm{orb}}^{2}},
\label{e1}\ 
\end{equation}
where $G$ is the gravitational constant and $a$ is the semimajor axis of 
the eccentric orbit. The Roche lobe radius for the WD $R_{\rm{L2}}$ at the periastron
can take the following form~\citep{1971ARA&A...9..183P}: 

\begin{equation}
\frac{R_{\rm{L2}}}{a \left(1-e \right)}=0.462\left(\frac{M_{\rm{WD}}}{M_{\rm{NS}}+M_{\rm{WD}}}\right)^{1/3} \ 
\label{e2}
\end{equation}
where $e$ is the eccentricity of the orbit. The WD radius $R_{\rm{WD}}$ can be expressed as~\citep{1988ApJ...332..193V}

\vspace{-12pt}
\begin{adjustwidth}{-\extralength}{0cm}
\centering 
\begin{equation}
R_{\rm{WD}}=0.0114 R_{\odot} \left[\left(\frac{M_{\rm{WD}}}{M_{\rm{Ch}}}\right)^{-2/3}-\left(\frac{M_{\rm{WD}}}{M_{\rm{Ch}}}\right)^{2/3}\right]^{1/2} \times \left[1+\frac{7}{2}\left(\frac{M_{\rm{WD}}}{M_{\rm{p}}}\right)^{-2/3}+\left(\frac{M_{\rm{WD}}}{M_{\rm{p}}}\right)^{-1}\right]^{-2/3}\ ,
\label{e3}
\end{equation}
\end{adjustwidth}
where $M_{\rm{Ch}}=1.44\ M_{\odot}$ is the Chandrasekhar mass limit and $M_{\rm{p}}=0.00057\ M_{\odot}$ {\citep{2004MNRAS.350..113M}}{.} 

\begin{figure}[H]

\includegraphics[height=8cm,width=10cm]{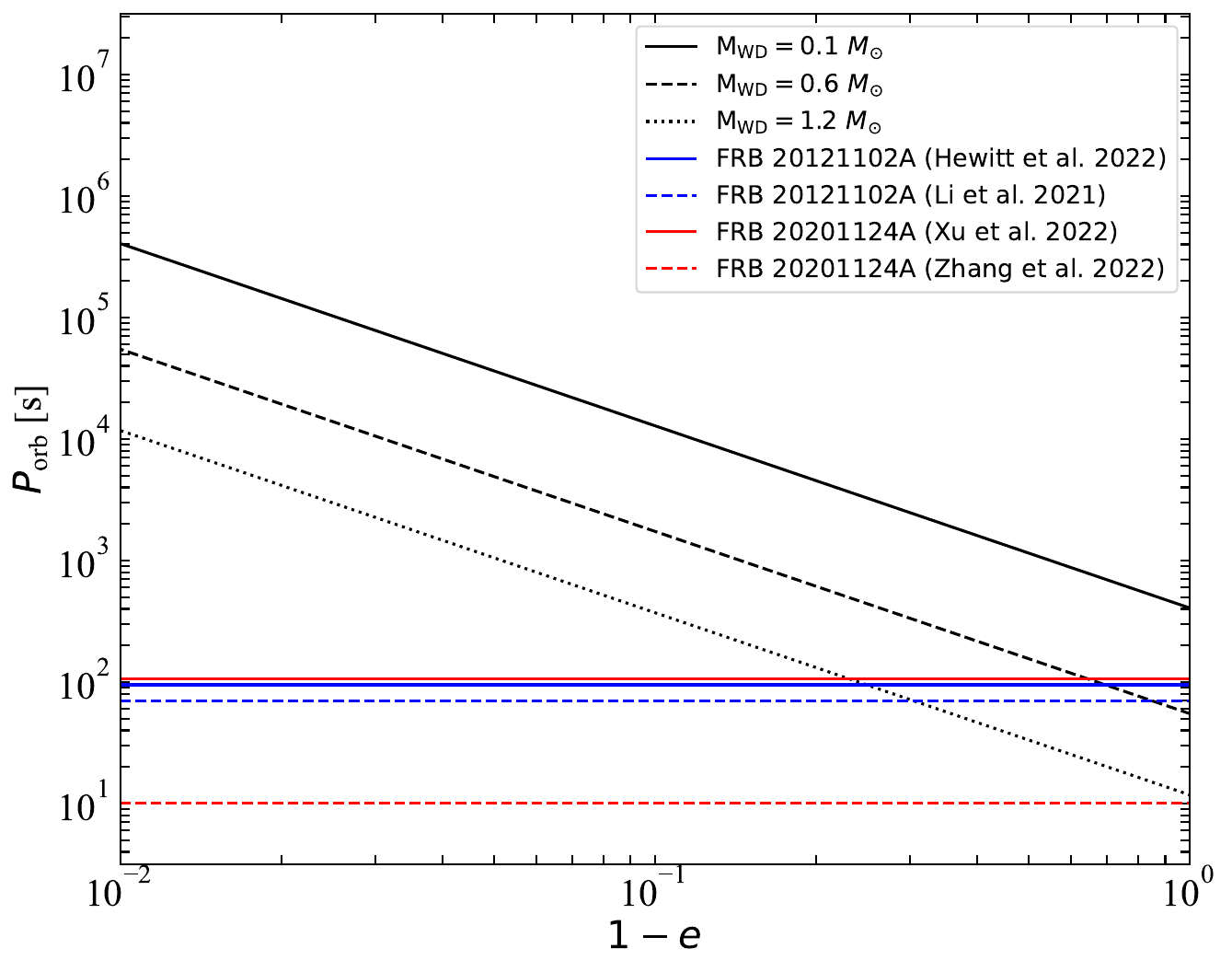}
\caption{Relationship between the orbital period $P_{\rm{orb}}$ and eccentricity
$e$ for the different WD masses $M_{\rm{WD}}$ (where $M_{\rm{NS}}=1.4\ M_{\odot}$) 
{when the WD fills its Roche lobe at periastron}.
The black lines represent difference WD masses, i.e., $0.1 \ M_{\odot}$ (solid line), $0.6 \ M_{\odot}$ (dashed line), and $1.2 \ M_{\odot}$ (dotted line). The blue lines correspond to the peak waiting times of FRB~20121102A reported by \mbox{\citet{Hewitt et al.(2022)}} (solid line) and \citet{Li et al.(2021)} (dashed line), respectively. The red lines represent the peak waiting times of FRB~20201124A reported by \citet{Xu et al.(2022)} (solid line) and \citet{Zhang et al.(2022)} (dashed line), respectively.
\label{fig1} }
\end{figure}

 At the periastron, it is assumed that the WD fills its Roche lobe, 
i.e., $R_{\rm{WD}}={R_{\rm{L2}}}$. The orbital period $P_{\rm{orb}}$ can be derived by 
Equations~(\ref{e1})--(\ref{e3}) after $M_{\rm{NS}}$, $M_{\rm{WD}}$, and $e$ are given.
The relationships between the orbital period $P_{\rm{orb}}$ and the eccentricity $e$ of the orbit for the distinct WD masses are shown in Figure~\ref{fig1}. The black lines represent different WD masses, i.e., $0.1 \ M_{\odot}$ (solid line), $0.6 \ M_{\odot}$ (dashed line), and $1.2 \ M_{\odot}$ (dotted line). The blue and red lines correspond to the peak waiting times of FRBs~20121102A and 20201124A, respectively, at the distinct activity epochs. It can be seen from Figure~\ref{fig1} that the orbital period $P_{\rm{orb}}$ of the NS-WD binary shortens with decreasing eccentricity $e$ for a constant WD mass.

Observations of the active repeating FRBs~20121102A and 20201124A reveal that the long-duration wait time peaks both shorten over time. In the NS-WD binary model, the WD mass remains quasi-constant for short-term (e.g.,~a few years) stable mass transfer processes, whereas the eccentricity $e$ of the NS-WD binary system can gradually decrease, as the dynamical mass transfer and gravitational radiation may cause orbital circularization. Thus, the orbital period $P_{\rm{orb}}$ can decrease gradually. 
As shown in Figure~\ref{fig1}, the orbital period $P_{\rm{orb}}$ varies from hundreds to tens of seconds as the eccentricity decreases when $M_{\rm{WD}} \gtrsim 0.6 \ M_{\odot}$, which is comparable to the variations in the long-duration peaks of the wait time distributions of FRBs~20121102A and 20201124A.

\subsection{Orbital Evolution of NS-WD Binary} \label{sec:evolution}

The orbital angular momentum of the NS-WD binary is provided by~\citep{1964PhRv..136.1224P}

\begin{equation}
J_{\rm orb}=M_{\rm NS}M_{\rm WD}\left[\frac{G a (1-e^{2})}{M_{\rm tot}}\right]^{1/2}\ ,
\label{Jorb}
\end{equation}
where $M_{\rm tot}=M_{\rm NS}+M_{\rm WD}$ is the total mass of the binary.
The orbital angular momentum evolution of a compact binary is dominated by the changes due to the gravitational wave radiation ($\dot{J}_{\rm GW}$), mass transfer ($\dot{J}_{\rm mt}$), and coupling between the spin of the (primary) accretor and the orbit ($\dot{J}_{\rm so}$) (e.g.,~\citep{2004MNRAS.350..113M}). The changes in the orbital angular momentum can be expressed as follows:
\begin{equation}
\dot{J}_{\rm orb}=\dot{J}_{\rm GW}+\dot{J}_{\rm mt}+\dot{J}_{\rm so}\ .
\label{dotJ}
\end{equation}
{Because} 
  the magnitude of the orbital angular velocity of eccentric binaries is a periodic function of time, it is unlikely for the stellar rotation rate to be synchronized with the orbital motion during all orbital phases~\citep{Sepinsky et al.(2009)}. Therefore, for NS-WD binaries with an eccentric orbit, we ignore the effect of the coupling between the spin of the NS and the orbit on the evolution of the orbital angular momentum.

We can deduce the time derivative of the orbital angular momentum $\dot{J}_{\rm orb}$ based on Equation~(\ref{Jorb}), i.e., 

\begin{equation}
\frac{\dot{J}_{\rm orb}}{J_{\rm orb}}=\frac{1}{2}\frac{\dot{a}}{a}+\frac{\dot{M}_{\rm NS}}{M_{\rm NS}}+\frac{\dot{M}_{\rm WD}}{M_{\rm WD}}-\frac{1}{2}\frac{\dot{M}_{\rm tot}}{{M}_{\rm tot}}-\frac{e \dot{e}}{1-e^{2}}\ ,
\label{dotJorb}
\end{equation}
where $\dot{M}_{\rm tot}=\dot{M}_{\rm NS}+\dot{M}_{\rm WD}$ and $\dot{M}_{\rm WD}<0$. 
The change of the orbital angular momentum caused by the gravitational wave radiation is provided by~\citep{1964PhRv..136.1224P}

\begin{equation}
\frac{\dot{J}_{\rm GW}}{J_{\rm orb}}=-\frac{32}{5} \frac{G^{3}}{c^{5}} \frac{M_{\rm NS} M_{\rm WD} M_{\rm tot}}{a^{4}} \frac{\left(1+\frac{7}{8}e^{2}\right)}{\left(1-e^{2}\right)^{5/2}}\ .
\label{Jgw}
\end{equation}
{When} 
 the NS-WD binary is a detached system, the gravitational wave radiation can shrink the orbital separation by dissipating the angular momentum of the system. With orbital decay due to gravitational wave radiation, the radius of the Roche lobe gradually decreases. Eventually, mass transfer occurs from the WD to the NS through the inner Lagrange point $L_{1}$ when the WD fills its Roche lobe at the periastron.  The angular momentum transferred from the WD to the neighborhood of the accreting NS along with the transferred matter can induce a change in the orbital angular momentum, which is provided by~\citep{1988ApJ...332..193V,van Haaften et al.(2012)}

\begin{equation}
\dot{J}_{\rm mt}=-\sqrt{G M_{\rm NS} R_{\rm h}} \dot{M}_{\rm NS}\ ,
\label{Jmt}
\end{equation}
where $R_{\rm h}$ is an equivalent radius of the accreted matter orbiting the accreting NS. 
This radius $R_{\rm h}$ can be written as~\citep{1988ApJ...332..193V}
\begin{equation}
\frac{R_{\rm h}}{a}=r_{\rm h}=0.0883-0.04858\ {\rm log}\ q+0.011489\ {\rm log^{2}} \ q+0.020475\ {\rm log^{3}}\ q\ ,
\end{equation}
{where $10^{-3}<q<1$. In the case of a fully developed accretion disk around the accreting NS, the angular momentum caused by the mass flow is canceled by a reverse flow~\citep{1975ApJ...198..383L,1988ApJ...332..193V,van Haaften et al.(2012)}, i.e., $R_{\rm h}=0$.}

On the other hand, as one of the most active repeating FRBs to date, the isotropic peak luminosities of FRB~20201124A are very high (e.g.,~from $5\times 10^{37} \ \rm{erg\ s^{-1}}$ to \mbox{$3\times 10^{40} \ \rm{erg\ s^{-1}}$};~\citep{Xu et al.(2022)}), indicating that the emission mechanism of this FRB source should be efficient. Based on the bursts reported by \citet{Xu et al.(2022)} and \citet{Zhang et al.(2022)}, the average accretion rates of two active episodes can be estimated to be $\sim$$ 1.4\times 10^{-4} \ M_{\odot}\ \rm{yr^{-1}}$ and $\sim$$ 5.2\times 10^{-5} \ M_{\odot}\ \rm{yr^{-1}}$, respectively, assuming that $L=0.1 \dot{M}_{\rm NS}c^{2}$. Thus, the outflows should be taken into account in the NS-WD binary with super-Eddington accretion rates ($\dot{M}_{\rm Edd} \sim 10^{-8} \ M_{\odot}\ \rm{yr^{-1}}$). 
We use the accretion efficiency $\epsilon$ to describe the fraction of the matter lost by the WD that is accreted by the NS, i.e., 

\begin{equation}
\dot{M}_{\rm NS}=-\epsilon \dot{M}_{\rm WD}\ 
\end{equation}
and

\begin{equation}
\dot{M}_{\rm tot}=\left(1-\epsilon \right)\dot{M}_{\rm WD}\ ,
\end{equation}
where $0 \le \epsilon \le 1$. If $\epsilon=1$, then all matter transferred from the WD is accreted by the NS, that is, the total mass of the system is conserved (i.e., $\dot{M}_{\rm NS}=-\dot{M}_{\rm WD}$ and $\dot{M}_{\rm tot}=0$). The only way to lose orbital angular momentum in this case is through gravitational wave radiation.
If $\epsilon<1$, then the binary gradually loses mass. In addition to the gravitational wave radiation, the system can lose angular momentum along with the ejected matter:

\begin{equation}
\frac{\dot{J}_{\rm loss}}{J_{\rm orb}}=\gamma \frac{\dot{M}_{\rm tot}}{M_{\rm tot}}\ .
\label{Jloss}
\end{equation}
In the particular case where the ejected matter carries the specific orbital angular momentum of the accretor, 
$\gamma=q$ (e.g.,~\citep{van Haaften et al.(2012)}).

Combining Equations~(\ref{dotJorb})--(\ref{Jloss}), the change of the semimajor axis of the eccentric NS-WD binary during the mass transfer process can be written as follows:
\begin{equation}
\frac{\dot a}{2a}=\frac{\dot{J}_{\rm GW}}{J_{\rm orb}}-\frac{\dot{M}_{\rm WD}}{M_{\rm WD}}\left[\left(1-\epsilon q\right)-\frac{\left(1-\epsilon \right)q}{1+q}\left(\gamma +\frac{1}{2} \right)-\epsilon \sqrt{\frac{\left(1+q\right)r_{\rm h}}{1-e^{2}}} \right]+\frac{e \dot{e}}{1-e^{2}}\ .
\label{da}
\end{equation}
{It} 
 is seen that the orbital separation can decay due to the gravitational wave radiation, mass loss, and orbital circularization.

\textls[-15]{At the periastron, the mass transfer from the WD to the NS occurs through the inner Lagrange point $L_{1}$ when the WD fills its Roche lobe. The radii of WD and its Roche lobe can change due to the mass loss of the WD. To ensure the stability of mass transfer, the change in the Roche lobe radius of the WD ${R_{\rm{L2}}}$ should exceed the change in its radius $R_{\rm WD}$,~i.e., }

\begin{equation}
{\dot{R}_{\rm L2}}\gtrsim \dot{R}_{\rm WD}\ .
\label{RR}
\end{equation}
{By} 
 dividing both sides of Equation~(\ref{RR}) by~${R_{\rm{L2}}}=R_{\rm WD}$ and $\dot{M}_{\rm WD}/M_{\rm WD}<0$, a well known criterion for the stability of mass transfer can be obtained (e.g.,~\citep{1985ibs..book...39W, 1988ApJ...332..193V}):

\begin{equation}
\zeta_{\rm L2}\lesssim \zeta_{\rm WD}\ 
\end{equation}
where $\zeta_{\rm L2}=\partial {\rm ln}  {R_{\rm{L2}}}/\partial {\rm ln} M_{\rm WD}$ is the response of the Roche lobe radius of the WD to the mass loss and $\zeta_{\rm WD}=\partial {\rm ln} R_{\rm WD}/\partial {\rm ln} M_{\rm WD}$ is the response of the WD radius to the mass loss. If the WD does not obey this criterion to start transferring matter via Roche lobe overflow, then the increase in WD radius after mass loss is larger than the effective expansion of its Roche lobe. In this case, the Roche lobe overflow tends to be unstable. 

\begin{figure}[H]
\begin{adjustwidth}{-\extralength}{0cm}
\centering 
\includegraphics[height=8cm,width=15cm]{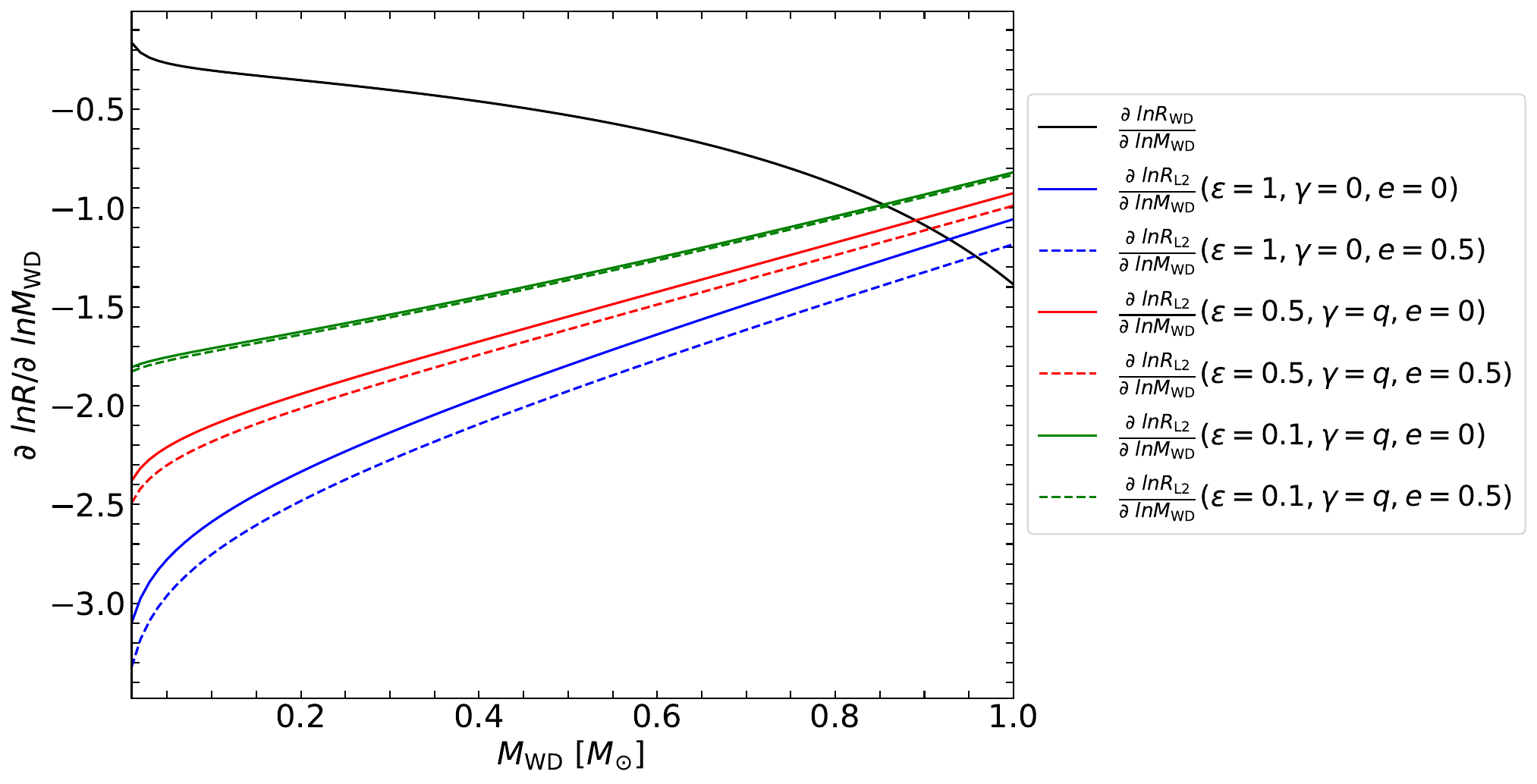}
\end{adjustwidth}
\caption{{Logarithmic} 
 changes of distinct radii due to mass transfer for different WD masses. The black line describes the logarithmic change in the WD radius via mass loss when $M_{\rm NS}=1.4\ M_{\odot}$. The solid and dashed lines represent the NS-WD binary with circular and eccentric orbits (e.g.,~$e=0.5$), respectively. The blue lines describe the logarithmic changes in the radius of the Roche lobe when the total mass of the system is conserved. The red and green lines describe the logarithmic changes in the Roche lobe radius when the total mass of the system gradually decreases, i.e., $\epsilon=0.5$ (red) and $\epsilon=0.1$ (green). Unstable mass transfer occurs when the black line describing the response of the WD radius falls below the colored lines describing the changes in the Roche lobe radius.
\label{fig2} }
\end{figure}

Based on Equation~(\ref{e3}), the logarithmic differentiation of WD radius $\zeta_{\rm WD}$ can be written~as
\begin{equation}
\zeta_{\rm WD}=\frac{\partial {\rm ln} R_{\rm WD}}{\partial {\rm ln} M_{\rm WD}}=-\frac{1}{3}\left[\frac{1+\left(\frac{M_{\rm WD}}{M_{\rm Ch}}\right)^{4/3}}{1-\left(\frac{M_{\rm WD}}{M_{\rm Ch}}\right)^{4/3}}\right]+\frac{2}{3}\left[\frac{1+\frac{7}{3}\left(\frac{M_{\rm WD}}{M_{\rm p}}\right)^{1/3}}{1+\frac{7}{2}\left(\frac{M_{\rm WD}}{M_{\rm p}}\right)^{1/3}+\left(\frac{M_{\rm WD}}{M_{\rm p}}\right)}\right]\ .
\label{dRwd}
\end{equation}
{The} 
 factor $\zeta_{\rm L2}$ can be derived by logarithmic differentiation of Equation~(\ref{e2}). We can deduce the critical mass ratio at which the unstable mass transfer occurs based on $\zeta_{\rm L2} \lesssim \zeta_{\rm WD}$. Figure~\ref{fig2} shows the logarithmic change in the radii of the WD and its Roche lobe due to mass transfer. In the NS-WD binary, we set the WD mass in the range of $0.01-1\ M_{\odot}$ and $M_{\rm NS}=1.4\ M_{\odot}$. {The black and colored lines describe the logarithmic change in the WD radius and Roche lobe radius, respectively, while the solid and dashed lines represent the NS-WD binary with circular and eccentric orbits (e.g.,~$e=0.5$), respectively. The blue lines describe the logarithmic change in the Roche lobe radius when the total mass of the system is conserved, i.e., $\epsilon=1$. The red and green lines describe the logarithmic changes in the Roche lobe radius when the total mass of the system gradually decreases,  i.e., $\epsilon=0.5$ (red) and $\epsilon=0.1$ (green). Unstable mass transfer occurs when the black line describing the response of the WD radius falls below the colored lines describing the changes in the Roche lobe radius.}
As shown in Figure~\ref{fig2}, mass transfer is dynamically unstable if $M_{\rm WD}>0.9\ M_{\odot}$,  corresponding to $q \gtrsim 2/3$ (e.g.,~\citep{King(2007),van Haaften et al.(2012)}). Moreover, the critical WD mass $M_{\rm c}$ in a NS-WD binary with an eccentric orbit is larger than that in a binary with a circular orbit. We can find that the critical WD mass decreases as the accretion efficiency $\epsilon$ decreases, i.e., $\epsilon=0.5$, $M_{\rm c} \sim 0.85\ M_{\odot}$, and $\epsilon=0.1$, $M_{\rm c} \sim 0.8\ M_{\odot}$. 

Based on the criterion $\zeta_{\rm L2}\lesssim \zeta_{\rm WD}$, the change of the eccentricity $e$ during the stable mass transfer process can be written as\vspace{-12pt}
\begin{adjustwidth}{-\extralength}{0cm}
\centering 
\begin{equation}
\frac{\dot{e}}{1+e} \lesssim \frac{2\dot{J}_{\rm GW}}{J_{\rm orb}}-\frac{2\dot{M}_{\rm WD}}{M_{\rm WD}}\left[\left(\frac{5}{6}-\epsilon q\right)-\frac{\left(1-\epsilon \right)q}{1+q}\left(\gamma+\frac{1}{3}\right)-\epsilon \sqrt{\frac{\left(1+q\right) r_{\rm h}}{1-e^{2}}}- \frac{\partial {\rm ln} R_{\rm WD}}{\partial {\rm ln} M_{\rm WD}} \right]\ .
\label{de}
\end{equation}
\end{adjustwidth}
{Therefore,} 
 we can combine Equations (\ref{da}) and (\ref{de}) to evaluate the orbital evolution of the NS-WD binary during the stable mass transfer process. 

\subsection{Mass Transfer Rate} \label{sec:rate}

For the NS-WD binary with an eccentric orbit, mass transfer from the WD to the NS occurs when the WD fills its Roche lobe at the periastron, i.e., $R_{\rm WD}=R_{\rm L2}$. We can define the overfill factor to indicate how much the donor overfills its Roche lobe by, i.e., 

\begin{equation}
\Delta=R_{\rm WD}-R_{\rm L2}\ .
\end{equation}
{Mass} 
 transfer occurs when $\Delta>0$. The adiabatic approximation of the mass transfer rate of the donor can be written as~\citep{2004MNRAS.350..113M} 

\begin{equation}
\dot{M}_{\rm WD}=-f\left(M_{\rm NS}, M_{\rm WD}, a, R_{\rm WD}\right) \Delta^{3}\ ,
\end{equation}
which increases monotonically with $\Delta$. As the system evolves, the mass transfer rate changes, leading to variations in the overfill factor. The change of the overfill factor is provided by~\citep{2004MNRAS.350..113M}

\begin{equation}
\frac{d \Delta}{dt}=R_{\rm WD}\left[\left(\zeta_{\rm WD}-\zeta_{\rm rL}\right)\frac{\dot{M}_{\rm WD}}{M_{\rm WD}}-\frac{\dot{a}}{a}\right]\ ,
\label{dDelta}
\end{equation}
where $\zeta_{\rm rL}=\partial {\rm ln} (R_{\rm L2}/a)/\partial {\rm ln} M_{\rm WD}$. A good approximation for the mass transfer rate can be obtained by setting the right-hand side of Equation~(\ref{dDelta}) to 0~\citep{2004MNRAS.350..113M,Sberna et al.(2021)}. Therefore, we obtain

\begin{equation}
\frac{\dot{M}_{\rm WD}}{M_{\rm WD}}=\frac{\dot{J}_{\rm GW}/J_{\rm orb}}{\frac{\left(\zeta_{\rm WD}-\zeta_{\rm rL}\right)}{2}+\left(1-\epsilon q\right)-\frac{\left(1-\epsilon\right)q}{1+q}\left(\gamma+\frac{1}{2}\right)-\epsilon \sqrt{\frac{\left(1+q\right) r_{\rm h}}{1-e^{2}}}}\ .
\label{dM}
\end{equation}

In the case of mass transfer in eccentric orbits, the evolution of the mass transfer rate has a Gaussian-like (or delta function) behavior, with the maximum mass transfer rate occurring at periastron~\citep{Sepinsky et al.(2007),Sepinsky et al.(2009),2016ApJ...825...71D}. Thus, we assume that Roche lobe outflow occurs only at periastron. The duration of the Roche lobe overflow can be evaluated as follows (e.g.,~\citep{2019ApJ...871L..17S}):

\begin{equation}
t_{\rm RLOF} \simeq R_{\rm p}/v_{\rm p}=\left(\frac{R_{\rm p}^{3}}{2GM_{\rm NS}}\right)^{1/2}\ ,
\label{tof}
\end{equation}
where $R_{\rm p}=a(1-e)$ is the distance between the binary components when the WD approaches periastron and $v_{\rm p}=(2GM_{\rm NS}/R_{\rm p})^{1/2}$ is the WD's orbital speed at $R_{\rm p}$. Combining Equations~(\ref{dM}) and (\ref{tof}), we can estimate the transferred mass from the WD during periastron passage. 

\subsection{Results} \label{sec:result}

We numerically integrate Equations (\ref{da}), (\ref{de}), and (\ref{dM}), starting from the onset of mass transfer. The short-term evolution (about one hundred years) of the NS-WD binary with an eccentric orbit is shown in Figure~\ref{fig3}. We assume that the original system consists of a $1.4 \ M_{\odot}$ NS and a $0.6 \ M_{\odot}$ WD with an eccentricity $e=0.3$. In this case, the initial orbital period of the binary is $\sim$$ 94$ s, comparable to the long-duration peaks of the wait time distributions of FRBs~20121102A~\citep{Hewitt et al.(2022)} and 20201124A~\citep{Xu et al.(2022)}. 

In Figure~\ref{fig3}, the different colored lines represent the evolution of the physical parameters under various accretion efficiencies $\epsilon$ and $\gamma$, i.e., $\epsilon=1, \gamma=0$ (blue), \mbox{$\epsilon=0.5, \gamma=q$} (red), and $\epsilon=0.1, \gamma=q$ (green). For $\epsilon=1$ and $\gamma=0$, the total mass of the system is conserved and all matter transferred from the WD is accreted by the NS. In this case, the only mechanism for losing orbital angular momentum is gravitational wave radiation. For the other two cases, the ejected matter provides an additional way to lose orbital angular momentum, resulting in a more significant reduction in both the orbital period and the eccentricity of the NS-WD binary. It can be seen from Figure~\ref{fig3} that the physical parameters of the NS-WD binary, including the orbital eccentricity $e$, semimajor axis $a$, and orbital period $P_{\rm orb}$ (panels a, b, and d of Figure~\ref{fig3}, respectively), can be significantly reduced during the mass transfer process; moreover, the variations of these parameters in the two cases of $\epsilon=0.5$ and $\epsilon=0.1$ are greater than those in the case of $\epsilon=1$. On the other hand, the WD mass loss rate $\dot{M}_{\rm WD}$ for the binary system with mass ejection is greater than that for the mass-conserving system (see panel c of Figure~\ref{fig3}). 

Although the orbital period of the NS-WD binary can be significantly reduced by the short-term Roche lobe overflow, the magnitude of the orbital period shrinkage is still smaller than that of the reductions in the long-duration waiting-time peaks of FRBs~20121102A and 20201124A (see panel d of Figure~\ref{fig3}). 

\begin{figure}[H]

\includegraphics[height=9cm,width=10cm]{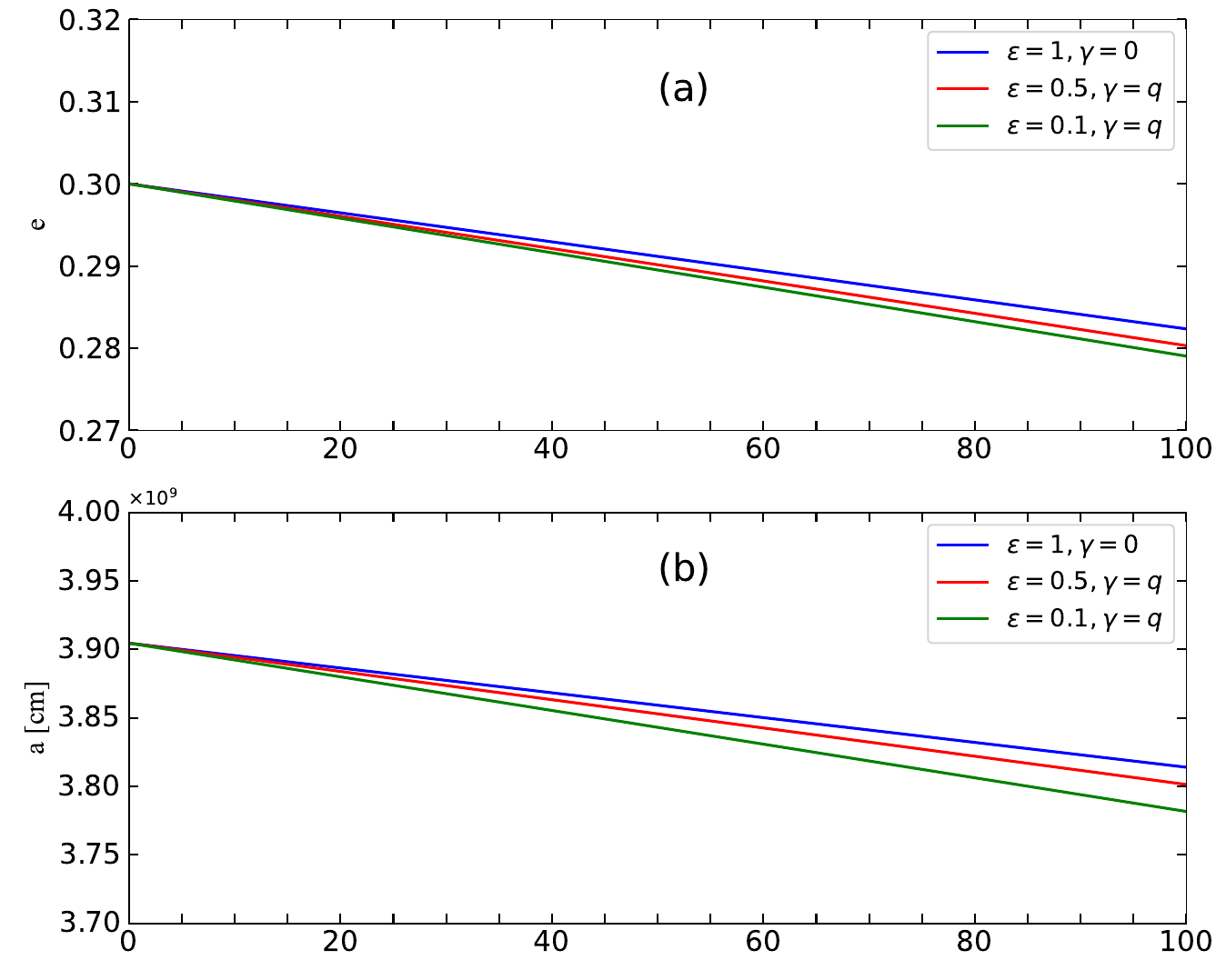}
\caption{\textit{Cont}.}
\label{fig:petit3leaks3D}
\end{figure}

\begin{figure}[H]\ContinuedFloat

\includegraphics[height=9cm,width=10cm]{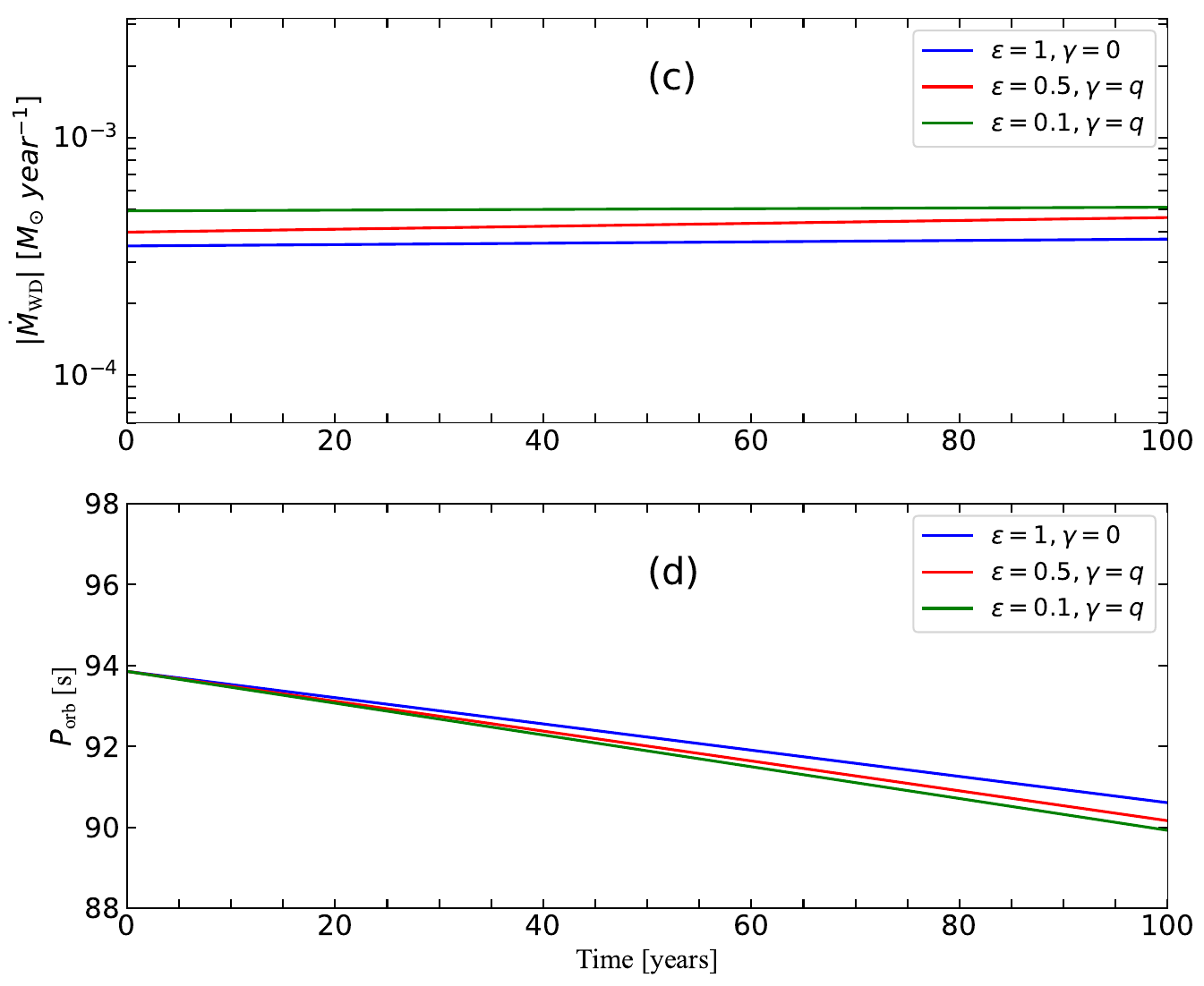}
\caption{{Evolution} 
 of the physical parameters of the NS-WD binary, including contributions from gravitational wave radiation, mass transfer, and mass ejection. The system with an eccentricity of $e=0.3$ consists of a $1.4\ M_{\odot}$ NS and a $0.6\ M_{\odot}$ WD. The panels show (\textbf{a}) the orbital eccentricity $e$, (\textbf{b}) the semimajor axis $a$, (\textbf{c}) the WD mass loss rate $\dot{M}_{\rm WD}$, and (\textbf{d}) the orbital period $P_{\rm orb}$. The different colored lines represent the various accretion efficiencies $\epsilon$ and $\gamma$, i.e., $\epsilon=1, \gamma=0$ (blue), $\epsilon=0.5, \gamma=q$ (red), and $\epsilon=0.1, \gamma=q$ (green).
\label{fig3} }
\end{figure}

\section{Unstable Mass Transfer} \label{sec:unstable}

{As shown in Figure~\ref{fig1}, in the case of $M_{\rm WD}=1.2\ M_{\odot}$ (dotted line), the range of the orbital periods of the NS-WD binaries with different eccentricities is consistent with the range of variations in the long-duration wait time peaks of both repeaters. The presence of a $1.2\ M_{\odot}$ WD suggests that the binary system undergoes an unstable mass transfer process (see Figure~\ref{fig2}), i.e., $\zeta_{\rm L2}> \zeta_{\rm WD}$, leading to evolution on dynamical time-scales.} In this case, the WD would expand dramatically over about tens of orbital periods~\citep{Bobrick et al.(2017)}, inducing a sharp increase in the mass transfer rate (e.g.,~$\dot{M}_{\rm WD} \gtrsim 100\ M_{\odot}\  \rm{yr^{-1}}$;~\citep{Bobrick et al.(2017)}). The NS is unable to accrete all the transferred materials, which subsequently pile up on the surface of the NS and begin to expand, eventually overfilling its Roche lobe. This results in the binary system becoming engulfed by a common envelope (CE). Due to the friction drag within the CE, the orbital energy of the embedded binary is reduced and deposited in the CE. If the total deposited energy can overcome the binding energy of the envelope, then orbital decay occurs, producing a binary with a very short orbital period following CE ejection. In contrast, if the CE fails to eject, then the inner binary coalesces into a single fast-rotating star (e.g.,~\citep{Han et al.(2020), 2024PrPNP.13404083C}).

The standard CE equation (known as the $\alpha$-mechanism) based on energy conservation can be written as~\citep{Webbink(1984),1988ApJ...329..764L}

\begin{equation}
\alpha_{\rm CE}\left(\frac{GM_{\rm 2}M_{\rm 1,f}}{2a_{\rm f}}-\frac{GM_{\rm 1,i}M_{\rm 2}}{2a_{\rm i}}\right)=\frac{GM_{\rm 1,i}\left(M_{\rm 1,i}-M_{\rm 1,f}\right)}{\lambda R_{\rm 1}}\ ,
\label{alpha}
\end{equation}
where $M_{\rm 1,i}$ and $M_{\rm 2}$ are the initial masses of the donor and the accretor, respectively, $M_{\rm 1,f}$ is the final mass of the donor that lost its envelope, $a_{\rm i}$ and $a_{\rm f}$ are the initial and final binary separations, respectively, and $R_{\rm 1}$ is the radius of the donor at the onset of CE. The left-hand term of Equation~(\ref{alpha}) is the energy $\Delta E_{\rm orb}$ released from orbital shrinkage, the right-hand term of Equation~(\ref{alpha}) is the binding energy, and $\lambda$ is a parameter describing the structure of the envelope. The CE ejection efficiency $\alpha_{\rm CE}$ is defined as the fraction of reduced orbital energy that is used in ejecting the CE. This $\alpha$-mechanism can be used to explain the formation and evolution of some close interacting binaries, including cataclysmic variables, low-mass X-ray binaries~\citep{Patterson(1984)}, and double-degenerate WD binaries~\citep{Webbink(1984)}. {However, some uncertainties exist in the $\alpha$-mechanism; for example, the exact values of $\alpha_{\rm CE}$ and $\lambda$ are hardly determined,} and the exact boundary between the stellar envelope and its core is hard to constrain (e.g.,~\citep{2024PrPNP.13404083C}).

If the angular momentum of the orbit is so large that the common envelope can easily be made to co-rotate with the orbit, then there are no drag forces anymore~\citep{Nelemans et al.(2000)}. In this case, it is assumed that the entire envelope is lost completely from the binary system~\citep{Nelemans et al.(2000),Nelemans et al.(2001)}. The mass loss process reduces the angular momentum of the system in a linear way, which can be written as follows:

\begin{equation}
\frac{J_{\rm i}-J_{\rm f}}{J_{\rm i}}=\gamma_{\rm CE} \frac{\Delta M}{M_{\rm tot}}\ 
\label{gamma}
\end{equation}
where $\Delta J=J_{\rm i}-J_{\rm f}$ is the decrease in the angular momentum due to CE ejection and $\Delta M$ is the envelope mass lost from the binary. This $\gamma$-mechanism is based on the angular momentum conservation, which was used in~\citep{Nelemans et al.(2000)} to reconstruct the evolutionary histories of three double-He WDs with $\gamma_{\rm CE}$ in the range of 1.4 to 1.7. The change in orbital period is provided by~\citep{Nelemans et al.(2000),Nelemans et al.(2001)}

\begin{equation}
\frac{P_{\rm f}}{P_{\rm i}}=\left(\frac{M_{\rm 1,f}M_{\rm 2,f}}{M_{\rm 1,i}M_{\rm 2,i}}\right)^{-3} \left(\frac{M_{\rm 1,f}+M_{\rm 2,f}}{M_{\rm 1,i}+M_{\rm 2,i}}\right)\left(1-\gamma_{\rm CE} \frac{\Delta M}{M_{\rm 1,i}+M_{\rm 2,i}}\right)^{3}\ ,
\label{Pf/Pi}
\end{equation}
\textls[-15]{where $P_{\rm i}$ and $P_{\rm f}$ are the orbital periods of the binary system before and after CE ejection, respectively, and $M_{\rm 2,i}$ and $M_{\rm 2,f}$ are the initial and final masses of the accretor,~respectively. }

Based on Equation~(\ref{Pf/Pi}), we can obtain the relationship between the ratio of the final to the initial orbital period $P_{\rm f}/P_{\rm i}$ and the mass loss $\Delta M_{\rm WD}$ due to CE ejection (Figure~\ref{fig4}). {The component masses of the NS-WD binary at the onset of CE are $M_{\rm NS}=1.4\ M_{\odot}$ and $M_{\rm WD}=1.2\ M_{\odot}$, respectively.} As shown in Figure~\ref{fig4}, different colored lines represent different $\gamma_{\rm CE}$, i.e., $\gamma_{\rm CE}=1$ (black), 2 (blue), 3 (green), and 4 (orange). {A larger value of $\gamma_{\rm CE}$ indicates that a greater fraction of the angular momentum of the initial binary is carried away by the ejected material~\citep{Nelemans et al.(2000)}.}
The red and purple dashed lines indicate the ratios of the peak waiting times in two different activity phases for FRBs~20121102A and 20201124A, respectively. 

\subsection{FRB~20121102A} \label{sec:121102}

To make the change in the orbital period of the NS-WD binary comparable to the variation in the wait time peaks of FRB~20121102A, the mass of the ejected material $\Delta M_{\rm WD}$ should range from $\sim$$ 0.5\ M_{\odot}$ to $\sim$$ 0.7\ M_{\odot}$ when $\gamma_{\rm CE}$ varies between 3 and 4 (see~the red dashed line in Figure~\ref{fig4}). After CE ejection, the remaining WD mass ($M_{\rm WD}\sim 0.5-0.7\ M_{\odot}$) falls below the critical mass ($M_{\rm c}\sim 0.9\ M_{\odot}$) and the mass transfer process of the NS-WD binary becomes stable. Subsequently, the eccentricity $e$ of both the orbit and the orbital period decrease as the orbital angular momentum decreases. 

The possible evolutionary history of FRB~20121102A is as follows. A massive WD ($M_{\rm WD}=1.2\ M_{\odot}$) may be tidally captured by the NS (e.g.,~\citep{Clark(1975)}), as large eccentricity $e$ of the orbit is required to interpret the initial orbital period of the system. After the formation of a detached NS-WD binary, the orbital separation can be reduced by the gravitational wave radiation, resulting in a steady decrease in the Roche lobe radius of the WD. In this stage, mass loss from the WD may occur through a tidally enhanced wind (e.g.,~\citep{1988MNRAS.231..823T}). The interaction between the wind and the magnetosphere of the NS could produce coherent radio emissions~\citep{Zhang(2017)}. When the WD fills its Roche lobe, dynamic unstable mass transfer occurs. A fraction of the transferred material approaches the NS surface, producing strong electromagnetic radiation by curvature radiation~\citep{Gu et al.(2020)}, whereas the majority of the transferred material cannot be accreted by the NS, leading to the formation of a CE. The angular momentum and the orbital period of the NS-WD binary both decrease due to the CE ejection. Therefore, the time interval between two adjacent bursts decreases as the orbital period shortens. After the CE ejection, the residual mass of the WD falls below the critical mass $M_{\rm c}$. Stable mass transfer can occur when the WD fills its Roche lobe at the periastron. Radio emissions can be produced when the accreted materials approach the surface of the NS. This burst phase could correspond to the burst storm observed by Arecibo~\citep{Hewitt et al.(2022)} and FAST~\citep{Li et al.(2021)}.
The orbital angular momentum of the NS-WD binary system is further reduced due to gravitational wave radiation and mass loss. When the eccentric orbit of the NS-WD binary system turns into a circular orbit, the mass transfer process always proceeds if the WD fills its Roche lobe. The FRB source may enter an extremely active phase, leading to a shorter time interval between two adjacent bursts.

In summary, the NS-WD binary model with an eccentric orbit can reconstruct the change in the burst activity of FRB~20121102A through the combined effects of CE ejection and Roche lobe overflow.

\subsection{FRB~20201124A} \label{sec:201124}

\begin{figure}[H]

\includegraphics[height=8cm,width=10cm]{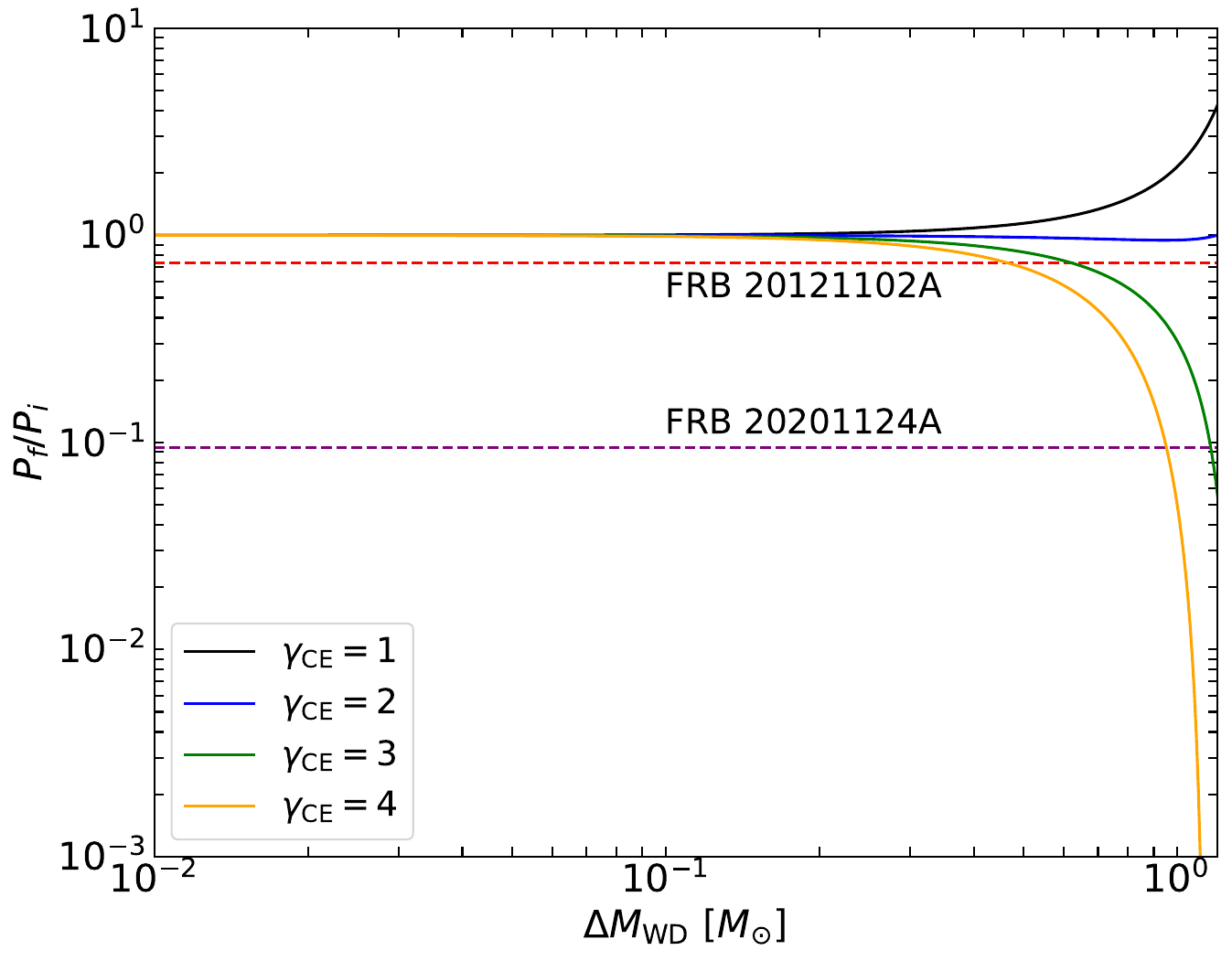}
\caption{Relationship between the orbital period ratio, defined as $P_{\rm f}/P_{\rm i}$, and the envelope mass lost from the binary $\Delta M$. Different colored lines represent different $\gamma_{\rm CE}$, i.e., $\gamma_{\rm CE}=1$ (black), 2 (blue), 3 (green), and 4 (orange). The initial physical parameters of the NS-WD binary system are $M_{\rm NS}=1.4\ M_{\odot}$ and $M_{\rm WD}=1.2\ M_{\odot}$. The red and purple dashed lines indicate the ratios of the wait time peaks in two different activity epochs for FRBs~20121102A and 20201124A, respectively.
\label{fig4} }
\end{figure}

For FRB~20201124A, as shown in Figure~\ref{fig4}, the mass of the ejected material exceeds $1\ M_{\odot}$ when $\gamma_{\rm CE}=3$--$4$, which can account for the changes in the waiting time peaks. After CE ejection, the binary system containing an NS and a naked O-Ne WD~\citep{Bobrick et al.(2017)} is in a circular orbit. Because the envelope of the WD is completely ejected, the mass transfer rate of the remaining component ($M_{\rm WD}<0.2\ M_{\odot}$, $\dot{M}_{\rm WD}\sim 10^{-7}\ M_{\odot}\ \rm{yr^{-1}}$;~\citep{Bobrick et al.(2017)}) is lower than that of the previous activity episodes, leading to a sharp decrease in the event rate of FRB~20201124A~\citep{Kirsten et al.(2024)}.

\citet{Nelemans et al.(2000)} considered two CE episodes to reconstruct the evolutionary histories of three double He WDs with $\gamma_{\rm CE}$ ranging from 1.4 to 1.7. In our model, the NS-WD binary system may also undergo two or more CE episodes with $\gamma_{\rm CE}$ ranging from 3 to 4, which would account for the variations in the wait time peaks of FRB~20201124A. After a CE ejection, the orbital separation decreases as the orbital angular momentum is reduced by the ejected material. Then, the Roche lobe radius of the WD would be reduced due to the gravitational wave radiation, causing the WD to fill its Roche lobe and triggering the next unstable mass transfer. In this scenario, the orbital period of the NS-WD binary becomes comparable to the waiting time peak after two or more CE episodes.

The orbital angular momentum of the evolved NS-WD binary is further reduced by gravitational wave radiation. Finally, the NS-WD binary merges. If the final remnant is a magnetar, then the NS-WD merged channel could produce one-off FRBs (e.g., FRB~20180924B;~\citep{2020ApJ...893....9Z}).

\section{Conclusions and Discussion}\label{sec:con}

In this paper, we {have revisited the} NS-WD binary model with an eccentric orbit to explain the observed variations in the long-duration wait time peaks of FRBs~20121102A and 20201124A. 
In our model, the decreases in the wait time peaks correspond to the decays in the orbital periods of the binary system. We consider two modes of mass transfer, namely, stable and unstable, to study the evolution of the orbital period of the binary system. Our main results are summarized as follows:

\begin{enumerate}

\item {In the case of stable mass transfer (e.g.,~$M_{\rm WD}=0.6\ M_{\odot}$), although the short-term Roche lobe overflow can significantly shorten the orbital period of the NS-WD binary, the magnitude of the orbital period shrinkage cannot fully account for the observed reductions in the long-duration wait time peaks of FRBs~20121102A and 20201124A (see Section~\ref{sec:stable}). }
\item {The CE ejections provide an additional and more efficient mechanism for angular momentum loss in the system with a massive WD (e.g.,~$M_{\rm WD}=1.2\ M_{\odot}$). By applying the $\gamma$-mechanism to the CE ejections, the variations in the orbital period of the NS-WD binary, which are comparable to the changes in the wait time peaks of two repeaters, can be reconstructed with $\gamma_{\rm CE}$ ranging between 3 and 4. Furthermore, our analysis also suggests distinct evolutionary pathways for the two sources; for FRB~20121102A, the binary likely undergoes a combination of CE ejection and Roche lobe overflow, while for FRB 20201124A the system may experience multiple CE ejections (see Figure~\ref{fig4}, Section~\ref{sec:unstable}).}
\item {For FRB~20121102A, the remnant WD mass falls below the critical mass $M_{\rm c}$ after the CE phase. Stable mass transfer occurs when the WD fills its Roche lobe at the periastron. For FRB~20201124A, the final binary system consisting of an NS and a stripped O-Ne WD should be in a circular orbit. Following the complete ejection of the WD envelope, the remaining mass transfer rate decreases significantly, leading to a sharp decrease in the event rate of FRB~20201124A (see Sections~\ref{sec:121102} and \ref{sec:201124}).}

\end{enumerate}

The observations of the two burst episodes for FRBs~20121102A and 20201124A are discontinuous~\citep{Li et al.(2021),Hewitt et al.(2022),Xu et al.(2022),Zhang et al.(2022)}, suggesting that the variations in the fitted wait time peaks across different activity epochs are also discontinuous. Thus, uncertainties remain in determining the evolutionary stage of the NS-WD binary based on the wait time peaks, such as estimating the WD mass before and after mass transfer.

Our evaluations indicate that the critical WD mass distinguishing stable from unstable mass transfer is $M_{\rm c}\sim 0.9\ M_{\odot}$. Furthermore, ejecting a sufficient amount of transferred material from the system could reduce the stability of mass transfer. Many studies have explored the stability of mass transfer in WD-NS systems. For example, \citet{1988ApJ...332..193V} showed that {if the transferred angular momentum is efficiently returned to the binary orbit by an accretion disk (i.e., $R_{\rm h}=0$), then the critical WD mass is $\sim$$ 0.6\ M_{\odot}$. In contrast, the critical WD mass without an accretion disk drops to around $0.3 \ M_{\odot}$.} Taking into account the isotropic re-emission mass transfer, \citet{van Haaften et al.(2012)} suggested a WD mass limit of $\sim$$ 0.8\ M_{\odot}$. Based on hydrodynamic simulations, {given that the angular momentum of the binary is lost by the disk wind, \citet{Bobrick et al.(2017)} reported a significantly lower critical WD mass of $\sim$$ 0.2\ M_{\odot}$.}

In addition to the variations in the long-duration wait time peaks, FRB~20201124A exhibits dramatic variations in the Faraday rotation measure on a day timescale, similar to FRB~20121102A (e.g.,~\citep{Michilli et al.(2018)}). These findings suggest the presence of a complex, dynamically evolving, and magnetized environment within the host galaxy~\citep{Xu et al.(2022)}. \citet{Wang et al.(2022)} proposed that FRB~20201124A may be produced by a binary system containing a magnetar and a Be star surrounded by a decretion disk. The orbital period is assumed to be 80 days, with orbital eccentricity $e=0.75$. The interactions between radio bursts, the decretion disk, and various magnetic field components can naturally explain the variable rotation measures.
In our model, the violent and unstable accretion process combined with the interactions between the radio bursts, the materials of the common envelope, and the magnetic field could explain the varying rotation measures~\citep{2022ApJ...937....5S}. After CE ejection, a non-evolving rotation measure can also be expected due to the vanishing of the CE~\citep{Jiang et al.(2022)}.

Moreover, a massive WD-NS merger could produce some gamma ray bursts (GRBs), e.g.,~GRB~211211A~\citep{Yang et al.(2022),Zhong et al.(2023)}. A millisecond magnetar is a likely outcome of such a merger if the magnetic field of the post-merger product is amplified through a mechanism such as an $\alpha$--$\Omega$ dynamo. Then, at least a subset of FRB~180924-like FRBs could be produced by the magnetar within the framework of flaring magnetars~\citep{2020ApJ...893....9Z}. On the other hand, due to their short orbital periods, NS-WD binaries are important sources of gravitational waves. The relationship between the gravitational wave frequency and the orbital period is given by $f=2/P_{\rm orb}$. Thus, the gravitational wave frequency of the closest NS-WD binary lies in the range of $10^{-2}$--$10^{-1}$ Hz,
which will potentially be observable by {the Laser Interferometer Space Antenna (LISA)} ($10^{-4}$--$10^{-1}$ Hz;~\citep[]{Amaro-Seoane et al.(2023)}) in the future.

Similar to FRBs~20121102A and 20201124A, the wait time distributions of the active repeaters FRBs~20220912A and 20240114A also exhibit bimodal structures~\citep{Zhang et al.(2023), Panda et al.(2024)}.
Double-peaked profiles of the wait time distributions seem to be a universal feature of active repeating FRBs. \citet{Xiao et al.(2024)} suggested that this asymmetric shape can be explained by the propagation effect in the magnetosphere of magnetars. \citet{Luo et al.(2025)} considered a rotation-modulated starquake model in magnetars to construct the wait time distribution. In future work, we aim to provide a detailed explanation of this bimodal structure based on the NS-WD binary system.

\funding{{This work was supported by the National Natural Science Foundation of China under grants 12221003 and 12433007.}} 

\institutionalreview{Not applicable.}

\informedconsent{Not applicable.}

\dataavailability{No new data were created or analyzed in this study.}

\acknowledgments{{We thank Wei-Min Gu and Tuan Yi for beneficial discussions, and thank the anonymous referees for helpful comments that improved the paper.}}

\conflictsofinterest{The author declares no conflicts of interest.} 

\abbreviations{Abbreviations}{
The following abbreviations are used in this manuscript:\\

\noindent 
\begin{tabular}{@{}ll}

CE & Common Envelope \\
CHIME & Canadian Hydrogen Intensity Mapping Experiment \\
FAST & Five-hundred-meter Aperture Spherical radio Telescope \\
FRB & Fast Radio Burst \\
GRB & Gamma Ray Burst \\
LISA & Laser Interferometer Space Antenna \\
NS & Neutron Star \\
WD & White Dwarf \\
WT & Waiting Time \\
\end{tabular}
}

\begin{adjustwidth}{-\extralength}{0cm}
\printendnotes[custom]

\reftitle{References}

\PublishersNote{}
\end{adjustwidth}

\begin{thebibliography}{999}

\bibitem[Cordes \& Chatterjee(2019)]{2019ARA&A..57..417C} Cordes, J.M.; Chatterjee, S. Fast Radio Bursts: An Extragalactic Enigma. {\em ARAA} {\bf 2019}, {\em 57}, 417--465.
\bibitem[Petroff et al.(2019)]{Petroff et al.(2019)} Petroff, E.; Hessels, J.W.T.; Lorimer, D.R. Fast radio bursts.
{\em AARv} {\bf 2019}, {\em 27}, 4.
\bibitem[Petroff et al.(2022)]{Petroff et al.(2022)} Petroff, E.; Hessels, J.W.T.; Lorimer, D.R. Fast radio bursts at the dawn of the 2020s. {\em AARv} {\bf 2022}, {\em 30}, 2.
\bibitem[Marcote et al.(2020)]{Marcote et al.(2020)} Marcote, B.; Nimmo, K.; Hessels, J.W.T.; Tendulkar, S.P; Bassa, C.G; Paragi, Z; Keimpema, A.; Bhardwaj, M.; Karuppusamy, R.; Kaspi, V.M.; et al. A repeating fast radio burst source localized to a nearby spiral galaxy. {\em Nature} {\bf 2020}, {\em 577}, 190--194.
\bibitem[Ryder et al.(2023)]{Ryder et al.(2023)} Ryder, S.D.; Bannister, K.W.; Bhandari, S., Deller, A.T.; Ekers, R.D.; Glowacki, M.; Gordon, A.C; Gourdji, K.; James, C.W.; \mbox{Kilpatrick, C.D.; et al.} A luminous fast radio burst that probes the Universe at redshift 1. 
{\em Science} {\bf 2023}, {\em 382}, 294--299.
\bibitem[Bochenek et al.(2020)]{Bochenek et al.(2020)} Bochenek, C.D.; Ravi, V.; Belov, K.V.; Hallinan, G.; Kocz, J.; Kulkarni, S.R.; McKenna, D.L. A fast radio burst associated with a Galactic magnetar.
{\em Nature} {\bf 2020}, {\em 587}, 59--62.
\bibitem[CHIME/FRB Collaboration et al.(2020b)]{CHIME/FRB Collaboration et al.(2020b)} Andersen, B.C.; Bandura, K.M.; Bhardwaj, M.; Bij, A.; Boyce, M.M.; Boyle, P.J.; Brar, C.; Cassanelli, T.; Chawla, P.; Chen, T.; et al. [The CHIME/FRB Collaboration]. A bright millisecond-duration radio burst from a Galactic magnetar. 
{\em Nature} {\bf 2020}, {\em 587}, 54--58.
\bibitem[Amiri et al.(2021)]{Amiri et al.(2021)} Amiri, M.; Andersen, B.C.; Bandura, K.; Berger, S.; Bhardwaj, M.; Boyce, M.M.; Boyle, P.J.; Brar, C.; Breitman, D.; \mbox{Cassanelli, T.; et al.} [The CHIME/FRB Collaboration]. The First CHIME/FRB Fast Radio Burst Catalog.
{\em Astrophys. J. Suppl. Ser.} {\bf 2021}, {\em 257}, 59.
\bibitem[Zhong et al.(2022)]{Zhong et al.(2022)} Zhong, S.Q.; Xie, W.J.; Deng, C.M.; Li, L.; Dai, Z.G.; Zhang, H.M. Can a Single Population Account for the Discriminant Properties in Fast Radio Bursts?
{\em Astrophys. J.} {\bf 2022}, {\em 926}, 206.
\bibitem[Chen et al.(2022b)]{Chen et al.(2022b)} Chen, H.Y.; Gu, W.M.; Sun, M.Y.; Yi, T. One-off and Repeating Fast Radio Bursts: A Statistical Analysis.
{\em Astrophys. J.} {\bf 2022}, {\em 939}, 27.
\bibitem[Cui et al.(2022)]{Cui et al.(2022)} Cui, X.H.; Zhang, C.M.; Li, D.; Zhang, J.W.; Peng, B.; Zhu, W.W.; Strom, R.; Wang, S.Q.; Wang, N.; Wu, Q.D.; et al. Luminosity distribution of fast radio bursts from CHIME/FRB Catalog 1 by means of the updated Macquart relation. 
{\em ApSS} {\bf 2022}, {\em 367}, 66.
\bibitem[Hessels et al.(2019)]{Hessels et al.(2019)} Hessels, J.W.T.; Spitler, L.G.; Seymour, A.D.; Cordes, J.M.; Michilli, D.; Lynch, R.S.; Gourdji, K.; Archibald, A.M.; Bassa, C.G.; Bower, G.C.; et al. FRB 121102 Bursts Show Complex Time-Frequency Structure.
{\em Astrophys. J. Lett.} {\bf 2019}, {\em 876}, L23.
\bibitem[Zhang(2020)]{Zhang(2020)} Zhang, B. The physical mechanisms of fast radio bursts. 
{\em Nature} {\bf 2020}, {\em 587}, 45--53.
\bibitem[Zhang(2023)]{Zhang(2023)} Zhang, B. The physics of fast radio bursts.
{\em RvMP} {\bf 2023}, {\em 95}, 035005.
\bibitem[Gopinath et al.(2024)]{Gopinath et al.(2024)} Gopinath, A.; Bassa, C.G.; Pleunis, Z.; Hessels, J.W.T.; Chawla, P.; Keane, E.F.; Kondratiev, V.; Michilli, D.; Nimmo, K. Propagation effects at low frequencies seen in the LOFAR long-term monitoring of the periodically active FRB 20180916B.
{\em Mon. Not. R. Astron. Soc.} {\bf 2024}, {\em 527}, 9872--9891.
\bibitem[CHIME/FRB Collaboration et al.(2020a)]{CHIME/FRB Collaboration et al.(2020a)} Amiri, M.; Andersen, B.C.; Bandura, K.M.; Bhardwaj, M.; Boyle, P.J.; Brar, C. et al. [The CHIME/FRB Collaboration]. Periodic activity from a fast radio burst source. 
{\em Nature} {\bf 2020}, {\em 582}, 351--355.
\bibitem[Chatterjee et al.(2017)]{Chatterjee et al.(2017)} Chatterjee, S.; Law, C.J.; Wharton, R.S.; Burke-Spolaor, S.; Hessels, J.W.T.; Bower, G.C.; Cordes, J.M.; Tendulkar, S.P.; Bassa, C.G.; Demorest, P.; et al. A direct localization of a fast radio burst and its host. 
{\em Nature} {\bf 2017}, {\em 541}, 58--61.
\bibitem[Marcote et al.(2017)]{Marcote et al.(2017)} Marcote, B.; Paragi, Z.; Hessels, J.W.T.; Keimpema, A.; van Langevelde, H.J.; Huang, Y.; Bassa, C.G.; Bogdanov, S.; Bower, G.C.; Burke-Spolaor, S.; et al. The Repeating Fast Radio Burst FRB 121102 as Seen on Milliarcsecond Angular Scales. 
{\em Astrophys. J. Lett.} {\bf 2017}, {\em 834}, L8.
\bibitem[Rajwade et al.(2020)]{Rajwade et al.(2020)}
Rajwade, K.M.; Mickaliger, M.B.; Stappers, B.W.; Morello, V.; Agarwal, D.; Bassa, C.G.; Breton, R.P.; Caleb, M.; Karastergiou, A.; Keane, E.F.; et al. Possible periodic activity in the repeating FRB 121102. 
{\em Mon. Not. R. Astron. Soc.} {\bf 2020}, {\em 495}, 3551--3558.
\bibitem[Cruces et al.(2021)]{Cruces et al.(2021)}
Cruces, M.; Spitler, L.G.; Scholz, P.; Lynch, R.; Seymour, A.; Hessels, J.W.T.; Gouiffés, C.; Hilmarsson, G.H.; Kramer, M.; Munjal, S. Repeating behaviour of FRB 121102: Periodicity, waiting times, and energy distribution. 
{\em Mon. Not. R. Astron. Soc.} {\bf 2020}, {\em 500}, 448--463.
\bibitem[Dai \& Zhong(2020)]{2020ApJ...895L...1D} Dai, Z.G.; Zhong, S.Q. Periodic Fast Radio Bursts as a Probe of Extragalactic Asteroid Belts. 
{\em Astrophys. J. Lett.} {\bf 2020}, {\em 895}, L1.
\bibitem[Gu et al.(2020)]{Gu et al.(2020)} Gu, W.M.; Yi, T.; Liu, T. A neutron star-white dwarf binary model for periodically active fast radio burst sources.
{\em Mon. Not. R. Astron. Soc.} {\bf 2020}, {\em 497},1543--1546.
\bibitem[Lyutikov et al.(2020)]{Lyutikov et al.(2020)} Lyutikov, M.; Barkov, M.V.; Giannios, D. FRB Periodicity: Mild Pulsars in Tight O/B-star Binaries.
{\em Astrophys. J. Lett.} {\bf 2020}, {\em 893}, L39.
\bibitem[Deng et al.(2021)]{Deng et al.(2021)} Deng, C.M.; Zhong, S.Q.; Dai, Z.G. An Accreting Stellar Binary Model for Active Periodic Fast Radio Bursts.
{\em Astrophys. J.} {\bf 2021}, {\em 922}, 98.
\bibitem[Chen et al.(2022a)]{Chen et al.(2022a)} Chen, H.Y.; Gu, W.M.; Fu, J.B.; Weng, S.S.; Wang, J.F.; Sun, M.Y. Repeating Ultraluminous X-Ray Bursts and Repeating Fast Radio Bursts: A Possible Association?
{\em Astrophys. J.} {\bf 2022}, {\em 937}, 9.
\bibitem[Lin et al.(2022)]{Lin et al.(2022)} Lin, Y.Q.; Chen, H.Y.; Gu, W.M.; Yi, T. Effects of Gravitational-wave Radiation of Eccentric Neutron Star-White Dwarf Binaries on the Periodic Activity of Fast Radio Burst Sources.
{\em Astrophys. J.} {\bf 2022}, {\em 929}, 114.
\bibitem[Beniamini et al.(2020)]{Beniamini et al.(2020)} Beniamini, P.;Wadiasingh, Z.; Metzger, B.D. Periodicity in recurrent fast radio bursts and the origin of ultralong period magnetars.
{\em Mon. Not. R. Astron. Soc.} {\bf 2020}, {\em 496},3390--3401.
\bibitem[Levin et al.(2020)]{Levin et al.(2020)} Levin, Y.; Beloborodov, A.M.; Bransgrove, A. Precessing Flaring Magnetar as a Source of Repeating FRB 180916.J0158+65.
{\em Astrophys. J. Lett.} {\bf 2020}, {\em 895}, L30.
\bibitem[Yang \& Zou(2020)]{2020ApJ...893L..31Y} Yang, H.; Zou, Y.C. Orbit-induced Spin Precession as a Possible Origin for Periodicity in Periodically Repeating Fast Radio Bursts.
{\em Astrophys. J. Lett.} {\bf 2020}, {\em 893}, L31.
\bibitem[Chen et al.(2021)]{Chen et al.(2021)} Chen, H.Y.; Gu, W.M.; Sun, M.; Liu, T.; Yi, T. Reconciling the 16.35-day Period of FRB 20180916B with Jet Precession.
{\em Astrophys. J.} {\bf 2021}, {\em 921}, 147.
\bibitem[Sridhar et al.(2021)]{Sridhar et al.(2021)} Sridhar, N.; Metzger, B.D.; Beniamini, P.; Margalit, B.; Renzo, M.; Sironi, L.; Kovlakas, K. Periodic Fast Radio Bursts from Luminous X-ray Binaries.
{\em Astrophys. J.} {\bf 2021}, {\em 917}, 13.
\bibitem[Li et al.(2021)]{Li et al.(2021)} Li, D.; Wang, P.; Zhu, W.W.; Zhang, B.; Zhang, X.X.; Duan, R.; Zhang, Y.K.; Feng, Y.; Tang, N.Y.; Chatterjee, S.; et al. A bimodal burst energy distribution of a repeating fast radio burst source. 
{\em Nature} {\bf 2021}, {\em 598}, 267--271.
\bibitem[Xu et al.(2022)]{Xu et al.(2022)} Xu, H.; Niu, J.R.; Chen, P.; Lee, K.J.; Zhu, W.W.; Dong, S.; Zhang, B.; Jiang, J.C.; Wang, B.J.; Xu, J.W.; et al. A fast radio burst source at a complex magnetized site in a barred galaxy. 
{\em Nature} {\bf 2022}, {\em 609}, 685--688.
\bibitem[Zhang et al.(2022)]{Zhang et al.(2022)} Zhang, Y.K.; Wang, P.; Feng, Y.; Zhang, B.; Li, D.; Tsai, C.W.; Niu, C.H.; Luo, R.; Yao, J.M.; Zhu, W.W.; et al. FAST Observations of an Extremely Active Episode of FRB 20201124A. II. Energy Distribution.
{\em RAA} {\bf 2022}, {\em 22}, 124002.
\bibitem[Lanman et al.(2022)]{Lanman et al.(2022)} Lanman, A.E.; Andersen, B.C.; Chawla, P.; Josephy, A.; Noble, G.; Kaspi, V.M.; Bandura, M.; Boyle, P.J.; Brar, C.; Breitman, D.; et al. A Sudden Period of High Activity from Repeating Fast Radio Burst 20201124A.
{\em Astrophys. J.} {\bf 2022}, {\em 927}, 59.
\bibitem[Niu et al.(2022)]{Niu et al.(2022)} Niu, J.R.; Zhu, W.W.; Zhang, B.; Yuan, M.; Zhou, D.J.; Zhang, Y.K.; Jiang, J.C.; Han, J.L.; Li, D.; Lee, K.J.; et al. FAST Observations of an Extremely Active Episode of FRB 20201124A. IV. Spin Period Search.
{\em RAA} {\bf 2022}, {\em 22}, 124004.
\bibitem[Hewitt et al.(2022)]{Hewitt et al.(2022)} Hewitt, D.M.; Snelders, M.P.; Hessels, J.W.T.; Nimmo, K.; Jahns, J.N.; Spitler, L.G.; Gourdji, K.; Hilmarsson, G.H.; Michilli, D.; Ould-Boukattine, O.S.; et al.
{\em Mon. Not. R. Astron. Soc.} {\bf 2022}, {\em 515}, 3577--3596.
\bibitem[Jiang et al.(2022)]{Jiang et al.(2022)} Jiang, J.C.; Wang, W.Y.; Xu, H.; Xu, J.W.; Zhang, C.F.; Wang, B.J.; Zhou, D.J.; Zhang, Y.K.; Niu, J.R.; Lee, K.J.; et al. FAST Observations of an Extremely Active Episode of FRB 20201124A. III. Polarimetry.
{\em RAA} {\bf 2022}, {\em 22}, 124003.
\bibitem[Zhou et al.(2022)]{Zhou et al.(2022)} Zhou, D.J.; Han, J.L.; Zhang, B.; Lee, K.J.; Zhu, W.W.; Li, D.; Jing, W.C.; Wang, W.Y.; Zhang, Y.K.; Jiang, J.C.; et al. FAST Observations of an Extremely Active Episode of FRB 20201124A: I. Burst Morphology.
{\em RAA} {\bf 2022}, {\em 22}, 124001.
\bibitem[King(2007)]{King(2007)} King, A.; Olsson, E.; Davies, M.B. A new type of long gamma-ray burst.
{\em Mon. Not. R. Astron. Soc.} {\bf 2007}, {\em 374}, L34--L36.
\bibitem[CHIME/FRB Collaboration et al.(2022)]{CHIME/FRB Collaboration et al.(2022)} Andersen, B.C.; Bandura, K.; Bhardwaj, M.;
Boyle, P.J.; Breitman, D.; Cassanelli, T.; Chatterjee, S.; Chawla, P.; Cliche, J.F.; \mbox{Cubranic, D.; et al.} Sub-second periodicity in a fast radio burst.
{\em Nature} {\bf 2022}, {\em 607}, 256--259.
\bibitem[Gu et al.(2016)]{Gu et al.(2016)} Gu, W.M.; Dong, Y.Z.; Liu, T.; Ma, R.Y.; Wang, J.F. A Neutron Star-White Dwarf Binary Model for Repeating Fast Radio Burst 121102.
{\em Astrophys. J. Lett.} {\bf 2016}, {\em 823}, L28.
\bibitem[Paczy{\'n}ski(1971)]{1971ARA&A...9..183P} Paczy{\'n}ski, B. Evolutionary Processes in Close Binary Systems.
{\em ARAA} {\bf 1971}, {\em 9}, 183.
\bibitem[Verbunt \& Rappaport(1988)]{1988ApJ...332..193V} Verbunt, F; Rappaport, S. Mass Transfer Instabilities Due to Angular Momentum Flows in Close Binaries.
{\em Astrophys. J.} {\bf 1988}, {\em 332},~193.
\bibitem[Marsh et al.(2004)]{2004MNRAS.350..113M} Marsh, T.R.; Nelemans, G.; Steeghs, D. Mass transfer between double white dwarfs. {\em Mon. Not. R. Astron. Soc.} {\bf 2004}, {\em 350}, 113--128.
\bibitem[Peters(1964)]{1964PhRv..136.1224P} Peters, P.C. Gravitational Radiation and the Motion of Two Point Masses.
{\em Phys. Rev.} {\bf 1964}, {\em 136}, 1224.
\bibitem[Sepinsky et al.(2009)]{Sepinsky et al.(2009)} Sepinsky, J.F.; Willems, B.; Kalogera, V.; Rasio, F.A. Interacting Binaries with Eccentric Orbits. II. Secular Orbital Evolution due to Non-conservative Mass Transfer. 
{\em Astrophys. J.} {\bf 2009}, {\em 702}, 1387.
\bibitem[van Haaften et al.(2012)]{van Haaften et al.(2012)} van Haaften, L.M.; Nelemans, G.; Voss, R.; Wood, M.A.; Kuijpers, J. The evolution of ultracompact X-ray binaries. 
{\em Astron. Astrophys.} {\bf 2012}, {\em 537}, A104.
\bibitem[Lubow \& Shu(1975)]{1975ApJ...198..383L} Lubow, S.H.; Shu, F.H. Gas dynamics of semidetached binaries.
{\em Astrophys. J.} {\bf 1975}, {\em 198}, 383.
\bibitem[Webbink(1985)]{1985ibs..book...39W} Webbink, R.F. Stellar evolution and binaries.
{\em Interact. Bin. Stars} {\bf 1985}, {39.} 

\bibitem[Sberna et al.(2021)]{Sberna et al.(2021)} Sberna, L.; Toubiana, A.; Miller, M.C. Golden Galactic Binaries for LISA: Mass-transferring White Dwarf Black Hole Binaries.
{\em Astrophys. J.} {\bf 2021}, {\em 908}, 1.
\bibitem[Sepinsky et al.(2007)]{Sepinsky et al.(2007)} Sepinsky, J.F.; Willems, B.; Kalogera, V.; Rasio, F.A. Interacting Binaries with Eccentric Orbits: Secular Orbital Evolution Due to Conservative Mass Transfer.
{\em Astrophys. J.} {\bf 2007}, {\em 667}, 1170.
\bibitem[Dosopoulou \& Kalogera(2016)]{2016ApJ...825...71D} Dosopoulou, F.; Kalogera, V. Orbital Evolution of Mass-transferring Eccentric Binary Systems. II. Secular Evolution. 
{\em Astrophys. J.} {\bf 2016}, {\em 825}, 71.
\bibitem[Shen(2019)]{2019ApJ...871L..17S} Shen, R.F. Fast, Ultraluminous X-Ray Bursts from Tidal Stripping of White Dwarfs by Intermediate-mass Black Holes.
{\em Astrophys. J. Lett.} {\bf 2019}, {\em 871}, L17.
\bibitem[Bobrick et al.(2017)]{Bobrick et al.(2017)} Bobrick, A.; Davies, M.B.; Church, R.P. Mass transfer in white dwarf-neutron star binaries.
{\em Mon. Not. R. Astron. Soc.} {\bf 2017}, {\em 467}, 3556--3575.
\bibitem[Han et al.(2020)]{Han et al.(2020)} Han, Z.W.; Ge, H.W.; Chen, X.F.; Chen, H.L. Binary Population Synthesis.
{\em RAA} {\bf 2020}, {\em 20}, 161.
\bibitem[Chen et al.(2024)]{2024PrPNP.13404083C} Chen, X.F.; Liu, Z.W.; Han, Z.W. Binary stars in the new millennium.
{\em PrPNP} {\bf 2024}, {\em 134}, 104083.
\bibitem[Livio \& Soker(1988)]{1988ApJ...329..764L} Livio, M.; Soker, N. The Common Envelope Phase in the Evolution of Binary Stars.
{\em Astrophys. J.} {\bf 1988}, {\em 329}, 764.
\bibitem[Webbink(1984)]{Webbink(1984)} Webbink, R.F. Double white dwarfs as progenitors of R Coronae Borealis stars and type I supernovae. 
{\em Astrophys. J. Suppl. Ser.} {\bf 1984}, {\em 277}, 355.
\bibitem[Patterson(1984)]{Patterson(1984)} Patterson, J. The evolution of cataclysmic and low-mass X-ray binaries.
{\em Astrophys. J. Suppl. Ser.} {\bf 1984}, {\em 54}, 443.
\bibitem[Nelemans et al.(2000)]{Nelemans et al.(2000)} Nelemans, G.; Verbunt, F.; Yungelson, L.R.; Portegies Zwart, S. Reconstructing the evolution of double helium white dwarfs: Envelope loss without spiral-in.
{\em Astron. Astrophys.} {\bf 2000}, {\em 360}, 1011--1018.
\bibitem[Nelemans et al.(2001)]{Nelemans et al.(2001)} Nelemans, G.; Yungelson, L.R.; Portegies Zwart, S.F.; Verbunt, F. Population synthesis for double white dwarfs . I. Close detached systems.
{\em Astron. Astrophys.} {\bf 2001}, {\em 365}, 491--507.
\bibitem[Clark(1975)]{Clark(1975)} Clark, G.W. X-ray binaries in globular clusters.
{\em Astrophys. J. Lett.} {\bf 1975}, {\em 199}, L143--L145.
\bibitem[Tout \& Eggleton(1988)]{1988MNRAS.231..823T} Tout, C.A; Eggleton, P.P. Tidal enhancement by a binary companion of stellar winds from cool giants.
{\em Mon. Not. R. Astron. Soc.} {\bf 1988}, {\em 231}, 823--831.
\bibitem[Zhang(2017)]{Zhang(2017)} Zhang, B. A {\textquotedblleft}Cosmic Comb{\textquotedblright} Model of Fast Radio Bursts.
{\em Astrophys. J. Lett.} {\bf 2017}, {\em 836}, L32.
\bibitem[Kirsten et al.(2024)]{Kirsten et al.(2024)} Kirsten, F.; Ould-Boukattine, O.S.; Herrmann, W.; Gawro{\'n}ski, M.P.; Hessels, J.W.T.; Lu,W.; Snelders, M.P.; Chawla, P.; Yang, J.; Blaauw, R.; et al. A link between repeating and non-repeating fast radio bursts through their energy distributions.
{\em Nat. Astron.} {\bf 2024}, {\em 8}, 337--346.
\bibitem[Zhong \& Dai(2020)]{2020ApJ...893....9Z} Zhong, S.Q; Dai, Z.G. Magnetars from Neutron Star-White Dwarf Mergers: Application to Fast Radio Bursts.
{\em Astrophys. J.} {\bf 2020}, {\em 893}, 9.
\bibitem[Michilli et al.(2018)]{Michilli et al.(2018)} Michilli, D.; Seymour, A.; Hessels, J.W.T.; Spitler, L.G.; Gajjar, V.; Archibald, A.M.; Bower, G.C.; Chatterjee, S.; Cordes, J.M.; Gourdji, K.; et al. An extreme magneto-ionic environment associated with the fast radio burst source FRB 121102.
{\em Nature} {\bf 2018}, {\em 553}, 182--185.
\bibitem[Wang et al.(2022)]{Wang et al.(2022)} Wang, F.Y.; Zhang, G.Q.; Dai, Z.G.; Cheng, K.S. Repeating fast radio burst 20201124A originates from a magnetar/Be star binary.
{\em Nat. Comm.} {\bf 2022}, {\em 13}, 4382.
\bibitem[Sridhar \& Metzger(2022)]{2022ApJ...937....5S} Sridhar, N.; Metzger, B.D. Radio Nebulae from Hyperaccreting X-Ray Binaries as Common-envelope Precursors and Persistent Counterparts of Fast Radio Bursts.
{\em Astrophys. J.} {\bf 2022}, {\em 937}, 5.
\bibitem[Yang et al.(2022)]{Yang et al.(2022)} Yang, J.; Ai, S.K.; Zhang, B.B.; Zhang, B.; Liu, Z.K.; Wang, X.Y.; Yang, Y.H.; Yin, Y.H.; Li, Y.; et al. A long-duration gamma-ray burst with a peculiar origin.
{\em Nature} {\bf 2022}, {\em 612}, 232--235.
\bibitem[Zhong et al.(2023)]{Zhong et al.(2023)} Zhong, S.Q.; Li, L.; Dai, Z.G. GRB 211211A: A Neutron Star-White Dwarf Merger?
{\em Astrophys. J. Lett.} {\bf 2023}, {\em 947}, L21.
\bibitem[Amaro-Seoane et al.(2023)]{Amaro-Seoane et al.(2023)} Amaro-Seoane, P.; Andrews, J.; Arca Sedda, M.; Askar, A.; Baghi, Q.; Balasov, R.; Bartos, S.S.; Bellovary, J.; Berry, C.P.L.; \mbox{Berti, E.; et al.} Astrophysics with the Laser Interferometer Space Antenna.
{\em Living Rev. Relativ.} {\bf 2023}, {\em 26}, 2.
\bibitem[Zhang et al.(2023)]{Zhang et al.(2023)} Zhang, Y.K.; Li, D.; Zhang, B.; Cao, S.; Feng, Y.; Wang, W.Y.; Qu, Y.H.; Niu, J.R.; Zhu, W.W.; Han, J.L.; et al. FAST Observations of FRB 20220912A: Burst Properties and Polarization Characteristics.
{\em Astrophys. J.} {\bf 2023}, {\em 955}, 142.
\bibitem[Panda et al.(2024)]{Panda et al.(2024)} Panda, U.; Roy, J.; Bhattacharyya, S.; Dudeja, C.; Kudale, S. Low-frequency, wideband study of an active repeater, FRB 20240114A, with the GMRT.
{\em arXiv} {\bf 2024}, {arXiv:2405.09749}.
\bibitem[Xiao et al.(2024)]{Xiao et al.(2024)} Xiao, D., Dai, Z.G.; Wu, X.F. The Propagation of Fast Radio Bursts in the Magnetosphere Shapes Their Waiting-time and Flux Distributions.
{\em Astrophys. J.} {\bf 2024}, {\em 962}, 35.
\bibitem[Luo et al.(2025)]{Luo et al.(2025)} Luo, J.W.; Niu, J.R.; Wang, W.Y.; Zhang, Y.K.; Zhou, D.J.; Xu, H.; Wang, P.; Niu, C.H.; Zhang, Z.H.; Zhang, S.; et al. Hyper-active repeating fast radio bursts from rotation modulated starquakes on magnetars.
{\em arXiv} {\bf 2025}, {arXiv:2502.16626}.

\end{thebibliography}
\end{document}